\documentclass[11pt,a4paper]{article}
\pdfoutput=1
\usepackage{jheppub}
\usepackage[english]{babel}
\usepackage{amsmath,amssymb,xcolor,mathtools}
\usepackage{MnSymbol}
\usepackage{hyperref}

\DeclareMathOperator{\tr}{Tr}
\DeclareMathOperator{\mathd}{d}
\DeclareMathOperator{\mathe}{e}
\DeclareMathOperator{\mathi}{i}
\DeclareMathOperator{\Aut}{Aut}

\newcommand{\lbracket}{\left\langle}
\newcommand{\rbracket}{\right\rangle}
\newcommand{\Z}{\mathcal{Z}}
\newcommand{\C}{\mathcal{C}}
\newcommand{\p}{\mathsf{p}}
\newcommand{\dvar}{\nu}
\newcommand{\eval}{\zeta}
\newcommand{\shift}{\hat{\mathsf{M}}_{i,\alpha}}
\newcommand{\newshift}{\hat{\mathsf{M}}_{i}}

\newcommand{\schur}{\mathsf{Schur}}
\newcommand{\jack}{\mathsf{Jack}}
\newcommand{\macdonald}{\mathsf{Macdonald}}
\newcommand{\constr}{\mathsf{U}}
\newcommand{\halfindex}{D^2\times_q S^1}

\usepackage{pgf}
\usepackage{tikz}
\usetikzlibrary{arrows,automata}
\usetikzlibrary{positioning}
\tikzset{
    box/.style={
           rectangle,
           rounded corners,
           draw=black, very thick,
           minimum height=2em,
           inner sep=2pt,
           text centered,
           },
    block/.style={
           rectangle,
           fill=gray!20,
           text width=7em,
           text centered,
           rounded corners,
           minimum height=2em
           },
}

\preprint{{\small{\textsf{UUITP-25/20}}}}

\title{On matrix models and their $q$-deformations}

\author{Luca Cassia,}
\author{Rebecca Lodin,}
\author{and Maxim Zabzine}

\affiliation{Department of Physics and Astronomy, Uppsala University,\\
Box 516, SE-75120 Uppsala, Sweden.}

\emailAdd{luca.cassia@physics.uu.se}
\emailAdd{rebecca.lodin@physics.uu.se}
\emailAdd{maxim.zabzine@physics.uu.se}

\abstract{Motivated by the BPS/CFT correspondence, we explore the similarities between the classical $\beta$-deformed Hermitean matrix model and the $q$-deformed matrix models associated to 3d $\mathcal{N}=2$ supersymmetric gauge theories on $\halfindex$ and $S_b^3$ by matching parameters of the theories. The novel results that we obtain are the correlators for the models, together with an additional result in the classical case consisting of the $W$-algebra representation of the generating function. Furthermore, we also obtain surprisingly simple expressions for the expectation values of characters which generalize previously known results.}

\arxivnumber{2007.10354}

\begin{document}
\maketitle
\flushbottom

\section{Introduction}

\begin{figure}[ht]
\caption{Schematic relation between the matrix models.}
\label{fig:scheme}
\begin{center}
\resizebox{\textwidth}{!}{%
\begin{tikzpicture}[->,>=stealth']
 \node[box, anchor=center] (D2S1)
 {%
 \begin{tabular}{c}
  \textbf{Gauge theory on $\halfindex$}\\
  {$(q,t)$-model}
 \end{tabular}
 };
 \node[box,
  right of=D2S1,
  node distance=7cm,
  anchor=center] (S3) 
 {%
 \begin{tabular}{c}
  \textbf{Gauge theory on $S^3_b$}\\
  {Modular double}
 \end{tabular}
 };
 \node[box,
  below of=D2S1,
  yshift=-2cm,
  anchor=center] (BETA) 
 {%
 \begin{tabular}{c}
  \textbf{Hermitean Matrix Model}\\
  {with $\beta$-deformation}
 \end{tabular}
 };
 \node[box,
  below of=BETA,
  yshift=-2cm,
  anchor=center] (HMM) 
 {%
 \begin{tabular}{c}
  \textbf{Hermitean Matrix Model}\\
  {$\int\mathrm{d}\Phi\,\mathrm{e}^{-\mathrm{Tr}V(\Phi)}$}
 \end{tabular}
 };
 \node[block,
  left of=D2S1,
  node distance=5cm,
 anchor=center] (Macdonald)
 {\textbf{Macdonald}};
 \node[block,
  left of=BETA,
  node distance=5cm,
 anchor=center] (Jack)
 {\textbf{Jack}};
 \node[block,
  left of=HMM,
  node distance=5cm,
 anchor=center] (Schur)
 {\textbf{Schur}};
 \path (D2S1) edge node[anchor=south]{$\times2$} (S3)
 (S3) edge[bend left] (BETA)
 (S3) edge[bend left] (HMM)
 (BETA) edge[bend right] node[anchor=left,right]{$q$-deformation} (D2S1)
 (D2S1) edge[bend right] node[anchor=left,left]{\begin{tabular}{c}Classical limit\\$t=q^\beta, q\to1$ \end{tabular}} (BETA)
 (HMM) edge[bend right] node[anchor=left,right]{$\beta$-deformation} (BETA)
 (BETA) edge[bend right] node[anchor=left,left]{\begin{tabular}{c}Schur limit\\$\beta\to1$ \end{tabular}} (HMM);
\end{tikzpicture}
}%
\end{center}
\end{figure}
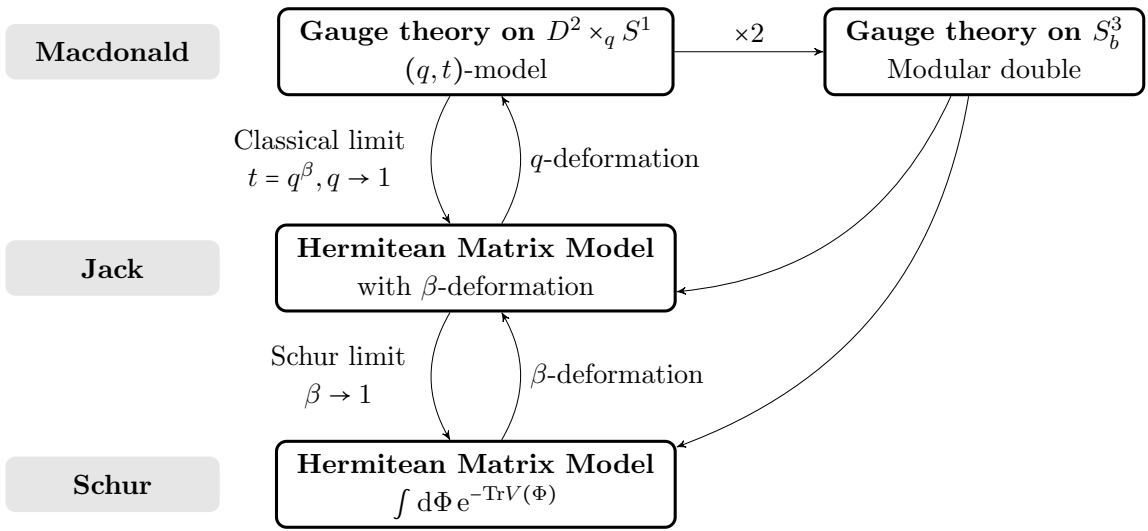

The BPS/CFT correspondence \cite{Nekrasov-BPS/CFT-1,Nekrasov-BPS/CFT-2} has provided a mapping between the exact values of partition functions and certain BPS observables for supersymmetric theories on the one hand, to conformal field theories in two dimensions on the other hand. The computation of partition functions and BPS observables has been aided through the programme called localization (as reviewed in \cite{Pestun:2016zxk}), where the evaluation of infinite dimensional integrals reduces to evaluation only at specific points. Such localization computations of partition functions sometimes result in the form of a finite dimensional matrix model, and it is such partition functions that will be of interest here.
\\
\\
In the light of this BPS/CFT correspondence, we derive and solve the Virasoro constraints that generating functions ($\tau$-function) for certain classical and ``quantum'' models satisfy. These constraints can be considered a type of Ward identities and were first studied in \cite{Mironov:1990im,Dijkgraaf:1990rs}. Here, the notion of ``quantum'' will be related to the introduction of a special class of $q$-deformations with respect to an un-deformed model. In the classical case, we will be considering the $\beta$-deformed Hermitean matrix model, in other words a one parameter deformation of the standard Hermitean matrix model \cite{Morozov:1994hh,Odake:1999un}. On the quantum side, we will explore the 3d $\mathcal{N}=2$ supersymmetric theory with $U(N)$ gauge group on both $\halfindex$ \cite{Beem:2012mb,Yoshida:2014ssa} and also $S_b^3$ \cite{Imamura:2011wg,Hama:2011ea} with one adjoint chiral and an arbitrary number of anti-chiral fundamental multiplets. The model on $\halfindex$ has also been referred to as the $(q,t)${\it -model} \cite{Lodin:2018lbz} where the parameters $q$ and $t$ are the two deformation parameters. In order to study the gauge theory on $S_b^3$ we will use a construction which has been called the {\it modular double} \cite{Nedelin:2016gwu}. The name refers to the picture of gluing two instances of $\halfindex$ to obtain $S^3_b$, something which is also mirrored in the algebraic structure at the level of partition functions. This modular double property of the $S_b^3$ partition function will be alluded to in Section~\ref{sec:quantum_models}. In the gauge theory examples these partition functions are expressed as matrix models, originating from a localization computation which we here simply assume the result of. Then, we both derive the constraint equations that these models satisfy explicitly and we also show how the resulting constraint can be solved in a recursive fashion. In other words we illustrate how any correlator of the theory can be determined using a finite number of steps of the recursion relation. In the case of the $\beta$-deformed Hermitean matrix model, we could in addition to the correlators also find the {\it $W$-algebra representation} of the generating functions. This is a representation in which the generating function is expressed through the action of a single operator acting on a simple function. Thus, the results which are novel here for the classical case are the correlators (presented in \eqref{eq:classical_correlators_p1} and \eqref{eq:classical_correlators_p2}) and the $W$-representations of generating functions (in \eqref{eq:WrepresentationP1} and \eqref{eq:WrepresentationP2}). This generalize the result of \cite{Morozov:2009xk,Itoyama:2017xid,Mironov:2017och} by introducing additional parameter dependence. In the case of the gauge theories on $\halfindex$ and $S_b^3$, the results are in terms of correlators (given in \eqref{eq:correlators_nf1} and \eqref{eq:correlators_nf2}), and they are extending the results of \cite{Cassia:2019sjk} by introducing another deformation parameter.
\\
\\
We also comment on the fact that averages of certain functions, when computed with respect to the measure of the partition function in question, take a particularly simple form. It is worth noting that this simplification is not expected a priori. These special functions are the Schur polynomials in the case of the standard Hermitean matrix model, Jack polynomials in the case of the $\beta$-deformed Hermitean matrix model and finally Macdonald polynomials in the case of the gauge theories on $\halfindex$ and $S_b^3$. The existence of such formulas has been referred to as the property of \textit{super-integrability} of the model \cite{Mironov:2017och,Mironov:2018ekq}. In the classical case we present formulas for the averages of Jack polynomials (in \eqref{eq:jackNf1} and \eqref{eq:jackNf2}), and in the quantum case we give the formulas for averages of Macdonald polynomials (in \eqref{eq:macdonaldNf1} and \eqref{eq:macdonaldNf2}) which improve and extend the results of \cite{Morozov:2018eiq}.
\\
\\
To further clarify the relation between the classical and the two quantum models we have in mind, we can illustrate the relations between the models as shown in Figure~\ref{fig:scheme}. Here we show the various deformations and limits to obtain one model from the other, together with the corresponding polynomial (whose average has a simple formula) for each model. Furthermore, we can also perform a matching between the parameters of the models as follows. The parameter $\beta$ of the classical Hermitean matrix model can be related to the mass $t$ of the adjoint chiral in the quantum model. The polynomial degree $\p$ of the potential $V$ in the classical model can be related to the number of fundamental anti-chiral fields $N_f$ in the quantum model. Then we can also match the coupling constants $a_k$ appearing in the classical potential $V$ with the masses of the fundamental anti-chiral fields $u_k$.
\\
\\
The outline of the paper is as follows. In Section~\ref{sec:classical_models} we begin with reviewing the basics of the simplest matrix model, the Hermitean 1-matrix model, and then show how the Virasoro constraints (or Ward identities) for the model are derived. We then show how to solve these constraints using a recursive procedure, where some of the results generalize previously known results. In Section~\ref{sec:quantum_models} we then move on to describe what could be considered quantum versions of the models outlined in Section~\ref{sec:classical_models}. Moreover, these models correspond to certain supersymmetric gauge theories and in particular 3d $\mathcal{N}=2$ theories with $U(N)$ gauge group on $\halfindex$ or $S_b^3$. Similarly to the previous section, we derive the $q$-Virasoro constraints which these models satisfy, and also show how to solve the constraints recursively to obtain novel results for the correlators of the models. A semi-classical expansion is also presented in order to match with the corresponding classical matrix model. Then in Section~\ref{sec:conclusion}, we conclude and suggest directions for further study. The details of special functions and of symmetric functions are left to Appendices~\ref{sec:special_functions} and \ref{sec:characters} respectively. In Appendix~\ref{sec:generators_relation}, we discuss the relation between the constraint operators and the generators of the $q$-Virasoro algebra and in Appendix~\ref{sec:asymptotic_analysis} we perform the analysis of the asymptotic behaviour and convergence of the $S_b^3$ partition function.

\section{Review of the classical models}
\label{sec:classical_models}

In this section we set the stage for the definition of the ``quantum'' matrix models by first reviewing the main features of the classical Hermitean matrix model (see \cite{Marino:2012zq,Eynard:2015aea} for an introduction to the topic). Here we present various well-known results by restating them in a way that makes it straightforward to match with the corresponding $q$-deformed case discussed in Section~\ref{sec:quantum_models}. {Moreover, we extend and improve upon the previously known formulas for the $W$-representation of the generating function. The main new results of our analysis are the formulas \eqref{eq:WrepresentationP1} and \eqref{eq:jackNf1} for the complex 1-matrix model, and \eqref{eq:WrepresentationP2} and \eqref{eq:jackNf2} for the Gaussian Hermitean matrix model.}

\subsection{Definitions}
\label{sec:classical_def}

Let us begin by recalling some details about classical matrix models. The simplest example of a matrix model is the Hermitean 1-matrix model (reviewed in \cite{Morozov:1994hh}), whose degrees of freedom are represented by the Hermitean $N\times N$ matrix $\Phi$.
The observables of the theory are the traces $\tr\Phi^s$ of the basic field and their expectation values can be neatly encoded into a generating function defined as
\begin{equation}
\label{eq:generatingfunctionabstract}
 \Z(\tau)=\int_{H_N}\mathd\Phi\,\mathe^{-\tr\,V(\Phi) + \sum_{s=1}^\infty\tau_s\tr\Phi^s}~.
\end{equation}
Here $V(\Phi)$ is a complex function (usually a polynomial) called the \textit{potential} while $\{\tau_s\}$ are conjugate variables to the traces and are usually referred to as the \textit{time variables} collectively denoted by $\tau$. The integral is over the domain $H_N$ which is taken to be the space of all $N\times N$ 
Hermitean matrices while the measure $\mathd{\Phi}$ is the standard Lebesgue measure on $H_N$ which is invariant under conjugation by unitary matrices.
{The generating function \eqref{eq:generatingfunctionabstract} is regarded as a formal power series in the times, however the coefficients of this expansion are integral functions and as such they must be convergent over the domain of integration and in some region of parameter space. For arbitrary potential $V$ there can be analytical issues with defining these integrals and for any specific choice one should perform an in-depth study.}
\\
\\
For a function $\mathcal{O}:H_N\to\mathbb{C}$, we define its (un-normalized) \textit{expectation value} or \textit{quantum average} as
\begin{equation}
 \langle\mathcal{O}\rangle = \int_{H_N} \mathd\Phi \,\mathcal{O}(\Phi)\,
 \mathe^{-\tr\,V(\Phi)}~,
\end{equation}
and for convenience we also define a time-dependent expectation value by inserting
$\mathcal{O}$ in the generating function as
\begin{equation}\label{eq:expectation_value}
 \langle\mathcal{O}\rangle_\tau = \int_{H_N} \mathd\Phi \,\mathcal{O}(\Phi)\,
 \mathe^{-\tr\,V(\Phi)+ \sum_{s=1}^\infty \tau_s \tr \Phi^s}~.
\end{equation}
Derivatives of the generating function $\Z(\tau)$ w.r.t. the times then compute the time-dependent expectation values of all possible single and multi-trace operators in the field $\Phi$. Sending all the times to zero then yields the corresponding time-independent average,
\begin{equation}
 \langle\mathcal{O}\rangle_{\tau=0}\equiv\langle\mathcal{O}\rangle~.
\end{equation}
As a function of the times $\{\tau_s\}$, $\Z(\tau)$ should be regarded as formal power series
which we can expand as
\begin{equation}
\begin{aligned}
 \Z(\tau) & = \lbracket \exp\left(\sum_{s=1}^\infty\tau_s\tr\Phi^s\right) \rbracket \\
 & = \sum_{n=0}^\infty \frac{1}{n!}\sum_{s_1=1}^\infty\dots\sum_{s_n=1}^\infty
 \langle\tr\Phi^{s_1}\dots\tr\Phi^{s_n}\rangle \tau_{s_1}\dots\tau_{s_n} \\
 & = \sum_{\rho}\frac{1}{|\Aut(\rho)|} c_\rho \prod_{a\in\rho}\tau_a
\end{aligned}
\end{equation}
where we rewrote the series as a sum over integer partitions $\rho$ of arbitrary size, and we also defined the \textit{correlation functions} $c_\rho$ as the expectation values of multi-trace operators whose powers are specified by the partition $\rho=(\rho_1,\dots,\rho_\ell)$, namely
\begin{equation}
 c_\rho := \langle\prod_{a\in\rho}\tr\Phi^a\rangle
 = \langle\tr\Phi^{\rho_1}\dots\tr\Phi^{\rho_\ell}\rangle~.
\end{equation}
The empty correlator $c_\emptyset$ is by definition equal to the \textit{partition function} $\Z=\Z(0)$.
\\
\\
If the potential is an invariant function under conjugation of the argument by a unitary matrix, then one can use the adjoint action of $U(N)$ over $H_N$ to diagonalize $\Phi$ and rewrite the generating function as an integral over the eigenvalues $\{\lambda_i\}$.
The Lebesgue measure splits as the product of a Vandermonde determinant $\Delta(\lambda)$, the flat measure $\prod_i\mathd\lambda_i$ over the space of eigenvalues and the Haar measure of $U(N)$,
\begin{equation}
 \mathd\Phi = \Delta(\lambda)^2 \prod_{i=1}^N \mathd \lambda_i \mathd U_{\mathrm{Haar}},
 \quad\quad\quad
 \Delta(\lambda) = \prod_{1\leq i < j\leq N} (\lambda_i-\lambda_j)~.
\end{equation}
Up to a constant overall factor, we can then write the integral as
\begin{equation}
\label{eq:hermitean_mm}
 \Z(\tau) = \int_{\mathbb{R}^N} \, \prod_{i = 1}^N \mathd \lambda_i
 \prod_{1 \leq i \neq j \leq N} (\lambda_i - \lambda_j)\,
 \mathe^{-\sum_{i=1}^N V(\lambda_i) +\sum_{s=1}^{\infty} \tau_s \sum_{i=1}^N \lambda_i^s}~.
\end{equation}
It is common at this point to introduce a 1-parameter deformation of the model by generalizing the usual Vandermonde term as follows (for details we refer to \cite{Odake:1999un})
\begin{equation}
 \Delta(\lambda) =
 \prod_{1 \leq i < j \leq N} (\lambda_i - \lambda_j) \xrightarrow{\beta\text{-deformation}} \prod_{1 \leq i < j \leq N} (\lambda_i - \lambda_j)^\beta
 = \Delta(\lambda)^\beta
\end{equation}
where $\beta$ is a positive integer number (which can be analytically continued to the complex plane).
Finally, the generating function becomes
\begin{equation}
\label{eq:beta_hermitean_mm_potential}
 \Z_{\beta}(\tau) = \int_{\mathbb{R}^N}\,\prod_{i=1}^N \mathd \lambda_i
 \prod_{1 \leq i \neq j \leq N} (\lambda_i - \lambda_j)^{\beta} 
 \mathe^{-\sum_{i=1}^N V(\lambda_i) + \sum_{s=1}^{\infty}  \tau_s \sum_{i=1}^N \lambda_i^s}~.
\end{equation}
In the limit $\beta\rightarrow 1$ one recovers the un-deformed model in \eqref{eq:hermitean_mm}. In the following we will always assume a $\beta$-deformation and therefore we will drop the label on the generating function.
\\
\\
For the purpose of our computations we are interested in the specific case of a potential function which is polynomial in the eigenvalues and takes the explicit form
\begin{equation}
 V(\lambda_i) = \sum_{k=1}^{\p} \frac{a_k}{k} \lambda_i^k~,
\end{equation}
which depends on the integer parameter $\p$ and on the complex numbers $a_k$ which can be regarded as inverse coupling constants\footnote{For $a_k=\delta_{k,2}$ one recovers the familiar Gaussian matrix model potential.}.
It is also worth mentioning that this kind of potential is special in the sense that we can obtain it via a constant shift of the time variables as
\begin{equation}
 \label{eq:shiftoftimes}
 \tau_s \mapsto \tau_s - a_s/s, \quad\quad\quad s=1,\dots,\p~.
\end{equation}
Assuming this potential, we will then for clarity introduce an index $\p$ on the generating function in \eqref{eq:beta_hermitean_mm_potential} and the expectation value \eqref{eq:expectation_value}, i.e. $\Z^{\p}(\tau)$ and $\langle \mathcal{O}\rangle_\tau^{\p}$.
Observe that while we assume that the dependence on the time variables $\{\tau_s\}$ is only formal,
after the shift \eqref{eq:shiftoftimes}, we need to carefully study the functional dependence of the generating function on the parameters $a_k$, and in particular we need to make sure that the integral in \eqref{eq:beta_hermitean_mm_potential} does indeed converge.
In the eigenvalue model of \eqref{eq:beta_hermitean_mm_potential} the contour of integration is taken to be the real domain $\mathbb{R}^N$ being the range of the eigenvalues of an Hermitean matrix in $H_N$, however when the potential $V(\lambda_i)$ is introduced one must modify the contour in such a way that the integral is still convergent, possibly complexifying the variables $\lambda_i$. For $\p\neq2$ for instance, if $\mathrm{Re}(a_\p)>0$ the integral is convergent and well-defined but only over half the real line\footnote{Observe that there are other choices of contour such that the integral is well-defined. For instance for $\p=3$ and $a_k=\delta_{k,3}$ the integral is a generalization of the Airy function for which one can define multiple contours going to infinity in the complex plane of $\lambda_i$ in different regions.}, i.e. for positive eigenvalues, while for $\p=2$ (and $\mathrm{Re}(a_2)>0$) the integral makes sense over the whole real $N$-dimensional space $\mathbb{R}^N$.
In general the Ward identities do not depend on a specific choice of contour (provided there are no additional boundary terms) and one can regard different contours as different \textit{branches} of the partition function, corresponding to different \textit{phases} of the theory.
\\
\\
For the special case $\p=1$ we also remark that the model we described has a close relative in the complex 1-matrix model \cite{MAKEENKO1991574, MORRIS1991703}
\begin{equation}
 \int_{M_N(\mathbb{C})}\mathd M\,\mathe^{-\tr\,V(M,M^\dagger) + \sum_{s=1}^{\infty} \tau_s \tr(MM^\dagger)^s}~,
\end{equation}
for a complex $N\times N$ matrix $M$ and its adjoint $M^\dagger$. Upon the change of variable to the Hermitean matrix $\Phi=MM^\dagger$, we can re-write the generating function as an integral over the positive eigenvalues $\lambda_i$. Taking the potential $V$ to be a quadratic function\footnote{While the potential $V(M,M^\dagger)$ is quadratic in $M$, in the eigenvalue variables $\lambda_i$ it becomes a polynomial of degree 1. This can be understood by noticing that if $M$ is diagonalizable with eigenvalues $\theta_i$, then the eigenvalues of $\Phi=MM^\dagger$ are $\lambda_i=|\theta_i|^2$.}, we can write the most general form of this model as that of the ($\beta$-deformed) Wishart-Laguerre eigenvalue model (reviewed for instance in \cite{Livan_2018})
\begin{equation}
\label{eq:WishartLaguerre}
 \Z^{\p=1}(\tau) = \int_{\mathbb{R}_{>0}^N}\,\prod_{i=1}^N \mathd \lambda_i
 \prod_{1 \leq i \neq j \leq N} (\lambda_i - \lambda_j)^{\beta} \prod_{i=1}^N \lambda_i^\dvar
 \mathe^{-a_1\sum_{i=1}^N \lambda_i + \sum_{s=1}^{\infty}  \tau_s \sum_{i=1}^N \lambda_i^s}~,
\end{equation}
where $\dvar$ is an additional parameter corresponding to the insertion of a determinant term of the form $(\det MM^\dagger)^\dvar$. The integral is convergent provided the power of the determinant satisfies
\begin{equation}
 \mathrm{Re}(\dvar)>-1~.
\end{equation}
Formally, we can reabsorb this term inside of the potential $V$ by writing it as a logarithmic interaction
\begin{equation}\label{eq:potential}
 V(\lambda_i) = -\delta_{\p,1} \dvar\log\lambda_i + \sum_{k=1}^{\p}\frac{a_k}{k} \lambda_i^k ~.
\end{equation}
Even though such determinant insertions are degenerate in the usual Hermitian matrix model because they make the Virasoro constraints ill-defined, in the case of $\p=1$, as we shall show in the next section, this insertion is allowed and indeed gives an additional 1-parameter deformation which has a direct counterpart in the quantum case.
\\
\\
Let us pause here to make a comment on conventions and notation. Since the matrix model
is built out of invariant functions w.r.t. the adjoint action of $U(N)$, we have that upon diagonalization and rewriting the generating function as an integral over the eigenvalues, there is a residual $S_N$ Weyl symmetry that permutes the variables $\{\lambda_i\}$. It is a well known fact that the ring of symmetric functions has a basis given by the \textit{power-sum} variables $\{p_s\}$ defined as
\begin{equation}
 p_s = \sum_{i=1}^N \lambda_i^s
\end{equation}
which are precisely the variables that couple to the times $\{\tau_s\}$ in the generating function. Derivatives with respect to the time $\tau_s$ correspond to the insertions of $p_s$. Another useful fact about symmetric function is that there exists an orthonormal basis provided by the Schur functions. The elements of this linear basis are symmetric polynomials labeled by integer partitions and are in 1-to-1 correspondence with the linear characters of $U(N)$. In the following we will often use that Schur polynomials can be expressed through power-sums, for example
\begin{equation*}
 \schur_{\{3\}} (p_k)= \frac{p_1^3}{6}+\frac{p_2 p_1}{2}+\frac{p_3}{3},
 \quad
 \schur_{\{2,1\}}(p_k) = \frac{p_1^3}{3}-\frac{p_3}{3},
 \quad
 \schur_{\{1,1,1\}}(p_k) = \frac{p_1^3}{6}-\frac{p_2 p_1}{2}+\frac{p_3}{3},
\end{equation*}
\begin{equation}
 \schur_{\{2\}}(p_k) = \frac{p_1^2}{2}+\frac{p_2}{2},
 \quad
 \schur_{\{1,1\}}(p_k) = \frac{p_1^2}{2}-\frac{p_2}{2},
\end{equation}
\begin{equation*}
 \schur_{\{1\}}(p_k) = p_1,
\end{equation*}
for all partitions of degree 3 and lower. We refer to Appendix~\ref{sec:characters} for more details on the subject. 
\\
\\
When one computes averages of Schur functions, these averages sometimes take a unexpectedly simple form as observed in \cite{Morozov:2018eiq}. For instance in the case of $\p=2$ and when $a_1=0$ we have the expression
\begin{equation} \label{eq:schurMorozov}
\left. \langle \schur_\rho(p_k) \rangle^{\p=2} \right|_{a_1=0} = \frac{1}{a_2^{|\rho|/2}} \frac{\schur_\rho (p_k = \delta_{k,2})}{\schur_\rho (p_k = \delta_{k,1})} \schur_\rho(p_k=N)
 \, c_{\emptyset}~.
\end{equation}
In what follows we will give examples of such averages.

\subsection{Virasoro constraints}
\label{sec:Virasoro_constraints}

It is often very useful in QFT to consider Ward identities for the path integral of the theory at hand. In the case of matrix models the QFT is 0-dimensional and the Ward identities have a very clear differential geometric interpretation. Let the partition function be described as the integral over a domain $X$ of the differential form $\Omega$.
If we consider an infinitesimal diffeomorphism generated by the vector field $\xi$ over $X$, then we can use the vector to deform infinitesimally the form $\Omega$ and then compute the integral of the variation as the Lie derivative, namely
\begin{equation}
\label{eq:deltaZ}
 \delta\Z = \int_X \mathcal{L}_\xi\Omega~.
\end{equation}
Since $\Omega$ is a top form on $X$, we can write the Lie derivative as an exact form
\begin{equation}
\label{eq:Liederivative}
 \mathcal{L}_\xi\Omega = \mathd\iota_\xi\Omega
\end{equation}
therefore its integral on $X$ can only receive contribution by evaluating the form $\iota_\xi\Omega$ at the boundary of $X$ (by Stokes theorem). Assuming that this form vanishes at the boundary, we get a non-trivial constraint equation corresponding to the fact that the variation $\delta\Z$ is identically zero.
\\
\\
In the case of the matrix models in Section~\ref{sec:classical_def} there is a natural family of vector fields given by
\begin{equation}
 \xi_n = \sum_{i=1}^N\lambda_i^{n + 1}\frac{\partial}{\partial\lambda_i},
\end{equation}
{which correspond to the generators of a Virasoro Lie algebra ${Vir}$ diagonally embedded into ${Vir}^N$, so that the vectors $\xi_n$ are invariant under permutations of the coordinates $\lambda_i$.}
The differential form $\Omega$ is the integrand in \eqref{eq:beta_hermitean_mm_potential} and the integration domain is $X=\mathbb{R}^N$.
Writing explicitly the top form as $\Omega=f(\lambda)\prod_{i=1}^N\mathd\lambda_i$, one can compute the total variation in \eqref{eq:Liederivative} as
\begin{equation}
\label{eq:totalvariation}
 \mathcal{L}_{\xi_n}\Omega = \sum_{i=1}^N\frac{\partial}{\partial\lambda_i}\left[\lambda_i^{n+1}f(\lambda)\right]\prod_{i=1}^N\mathd\lambda_i
\end{equation}
which upon integration together with the notation in \eqref{eq:expectation_value} leads to the Ward identity
\begin{equation}
\label{eq:virasoro_constraints_vev}
 \left\langle
  \beta \sum_{i, j=1}^N \sum_{k = 0}^n \lambda_i^k \lambda_j^{n - k} + (1 - \beta)
  (n + 1) \sum_{i=1}^N \lambda^n_i + \sum_{s>0} s \tau_s \sum_{i=1}^N \lambda^{s
  + n}_i -\sum_{k=1}^{\p} a_k \sum_{i=1}^N \lambda_i^{k+n}+\dvar\delta_{\p,1} \sum_{i=1}^N \lambda^n_i  \right\rangle^{\p}_{\tau} = 0~,
\end{equation}
These constraint equations are called the \textit{Virasoro constraints}. As mentioned above, the determinant insertion depending on the parameter $\dvar$ is only allowed in the case of $\p =1$. What one does at this point is to rewrite these expectation values as derivatives in times using the identity
\begin{equation}
 \lbracket \sum_{i_1=1}^N \dots\sum_{i_k=1}^N \lambda_{i_1}^{s_1}
 \dots \lambda_{i_k}^{s_k}\rbracket_\tau^\p
 = \frac{\partial}{\partial\tau_{s_1}}\dots\frac{\partial}{\partial\tau_{s_k}}\Z^\p(\tau)
\end{equation}
whenever all the powers $s_1,\dots,s_k$ are non-negative (if some of the $s_l=0$ we just substitute the corresponding sum with multiplication by $N$). This can be done for all $n\geq-1$
if $\dvar=0$, but for $\dvar\neq0$ the $n=-1$ constraint cannot be rewritten as a partial differential equation. The final form of the Virasoro constraints is then
\begin{equation}
\label{eq:virasoro_constraints}
 \underbrace{\left(\sum_{k=1}^{\p} a_k \frac{\partial}{\partial\tau_{k+n}}
 {+a_1 N \delta_{n,-1}}
 -\dvar\delta_{\p,1} \left(\frac{\partial}{\partial \tau_n}+\delta_{n,0}N\right) -L_n\right)}_{\constr_n}\Z^{\p}(\tau) =0
\end{equation}
where $\constr_n$ is the differential operator that implements the $n$-th constraint and the operators $L_n$ are the standard generators of the Virasoro algebra\footnote{Observe that, for $\dvar=0$, the operators $\constr_n$ can be obtained from the $L_n$ via the formal shift \eqref{eq:shiftoftimes} hence they satisfy the same Virasoro algebra.} defined as
\begin{equation}
\label{eq:virasoro_generators}
\begin{split}
& L_{n>0} = 2\beta N \frac{\partial}{\partial \tau_n} + \beta\sum_{a+b=n} \frac{\partial^2}{\partial \tau_a \partial \tau_{b}}+(1 - \beta) (n + 1) \frac{\partial}{\partial \tau_n} + \sum_{s>0} s \tau_s \frac{\partial}{\partial \tau_{s + n}} \\
& L_0 = \beta N^2 + (1-\beta)N+\sum_{s>0} s \tau_s \frac{\partial}{\partial \tau_{s }} \\
& L_{-1} = N \tau_1 +\sum_{s >0} s \tau_s \frac{\partial}{\partial \tau_{s -1}}  ~.
\end{split}
\end{equation}
By construction then, \eqref{eq:virasoro_constraints} states that the generating function $\Z^{\p}(\tau)$ is in the common kernel of all such operators $\constr_n$. In the following sections we study the properties of this kernel.

\subsection{Solving the constraints}

A legitimate question one might ask at this point is how strong are the Virasoro constraints in \eqref{eq:virasoro_constraints}. Can they be used to determine the generating function $\Z^{\p}(\tau)$ and if so, what is the degeneracy of the solution? The answer to these questions was found in \cite{Morozov:2009xk} via a $W$-algebra representation, which states that the solution is essentially unique if $\p=1,2$ while for $\p\geq3$ there is a degeneracy in the space of solutions which allows to determine $\Z^{\p}(\tau)$ only when additional information on the correlation functions is provided \cite{Cordova:2016jlu}. More recently in \cite{Morozov_2019} the solution for $\p=1,2$ was also found in the $\beta$-deformed model.
We will now review the details of the derivation of the solution to the constraints and the issues one encounters when such a unique solution does not exists.

\subsubsection{$\p = 1$}

The case $\p=1$ is the only one that admits a determinant insertion as discussed in \eqref{eq:WishartLaguerre}, thus in what follows we will always assume dependence on $\dvar$ in the case of $\p=1$. The $n=-1$ constraint in \eqref{eq:virasoro_constraints} is not well defined as a differential equation as it contains both
negative powers of the $\{\lambda_i\}$ as well as additional boundary terms. For these reasons we restrict ourselves to consider only Virasoro constraints for $n\geq0$
\begin{equation}
\label{eq:semiclassicalvirasoroNf1}
\begin{aligned}
 a_1\partial_{n+1} \Z^{\p=1}(\tau) =
 \bigg[ \beta \sum_{a+b=n} \partial_a\partial_b 
 & + \left((1-\beta)(n+1)+\dvar + 2\beta N \right) \partial_n + \\
 & + \sum_{s=1}^\infty s \tau_s \partial_{s+n}
 + \delta_{n,0}N\left(\dvar + \beta(N-1)+1\right)
 \bigg] \Z^{\p=1}(\tau)~,
\end{aligned}
\end{equation}
where from now on we use $ \partial_n = \frac{\partial}{\partial \tau_n}$ to ease notation. To obtain the solution we re-sum all constraints to construct the operator
\begin{equation}
\label{eq:p1classicalconstr}
 \constr = \sum_{n=0}^\infty (n+1)\tau_{n+1}\constr_{n} = a_1 D - W_{-1}
\end{equation}
which we have rewritten as the difference of two operators: $D=\sum_{s=1}^\infty s\tau_{s}\partial_{s}$ is the dilatation operator and
\begin{equation}
\begin{aligned}
 W_{-1} = & \beta \sum_{n,m=1}^\infty (n+m+1) \tau_{n+m+1} \partial_n\partial_m
 + \sum_{n,m=1}^\infty n m \tau_n \tau_m \partial_{n+m-1} + \\
 & + \tau_1 N (\dvar+\beta(N-1)+1)
 + \sum_{n=1}^\infty\left(\dvar+(1-\beta)(n+1)+2\beta N\right)(n+1)\tau_{n+1}\partial_{n}
\end{aligned}
\end{equation}
is a ``shifted'' $W$-algebra generator also called \textit{cut-and-join operator}. Here the word shifted refers to the fact that $W_{-1}$ is of degree 1 with respect to the grading introduced by the operator $D$ (i.e. $[D,W_{-1}]=W_{-1}$).
\\
\\
Let us analyze the properties of these operators. First we remark that they are linear operators acting on the infinite dimensional vector space underlying the commutative ring $\mathbb{C}[[\tau_1,\tau_2,\dots]]$. This vector space has a natural basis over $\mathbb{C}$ given by the monomials, i.e. products of times labeled by integer partitions $\rho$
\begin{equation}
 \prod_{a\in\rho}\tau_a ~.
\end{equation}
The ordering of the basis of the vector space is the one induced by the ordering on integer partitions, namely ordering by degree and lexicographic ordering between partitions of equal degree
\begin{equation}
 \emptyset < \{1\} < \{1,1\} < \{2\} < \{1,1,1\} < \{2,1\} < \{3\} < \dots ~.
\end{equation}
With these conventions in place one can show that $\constr$ is triangular and that $D$ is its diagonal part while $W_{-1}$ is its off-diagonal part. More precisely, one finds that $\constr$ is triangular also with respect to the weaker partial order induced by the monomial degree only (partition size).
Moreover, we have that $D$ acts on monomials as multiplication by the degree of the corresponding partition, 
\begin{equation}
 D\prod_{a\in\rho}\tau_a = \big(\sum_{a\in\rho}a\big)\prod_{a\in\rho}\tau_a
 = |\rho|\,\prod_{a\in\rho}\tau_a
\end{equation}
so that $\det\constr=\det D=0$ because there is one zero-eigenvalue corresponding to the empty partition. However, since all other eigenvalues are non-zero, the kernel of $\constr$ is exactly 1-dimensional. This means that the generating function $\Z^{\p=1}(\tau)$, regarded as a vector in this space, is uniquely defined up to a constant multiplicative factor which corresponds to the normalization of the trivial correlator $c_{\emptyset}$.
\\
\\
The full solution can be derived by recursively solving the equation
\begin{equation}
\label{eq:semiclassicalWvirasoroNf1}
 a_1 D \Z_{(d)}^{\p=1}(\tau) = W_{-1} \Z_{(d-1)}^{\p=1}(\tau),
 \quad\quad\quad
 \Z_{(d)}^{\p=1}(\tau) = \sum_{\rho\vdash d}\frac{1}{|\Aut(\rho)|}
 c_\rho \prod_{a\in\rho}\tau_a
\end{equation}
where $\rho \vdash d$ denotes that $\rho$ is an integer partition of $d$ with $d$ being the degree in times. Then, using the fact that $W_{-1}$ is of degree 1, we can write
\begin{equation}
\label{eq:WrepresentationP1}
 \Z^{\p=1}(\tau) = \sum_{d=0}^\infty \frac{W_{-1}^{d}}{a_1^{d}d!} \cdot c_{\emptyset}
 = \exp\left(\frac{W_{-1}}{a_1}\right)\cdot c_{\emptyset}~.
\end{equation}
A full solution of the $\p=1$ model, up to degree 3 is given by the correlators
\begin{equation}
\label{eq:classical_correlators_p1}
\begin{aligned}
c_{\{3\}} & = \frac{{N} (\dvar +\beta  ({N}-1)+1) \left(\dvar ^2+5 \dvar +5 \beta  \dvar  ({N}-1)+\beta  ({N}-1) (\beta  (5 {N}-6)+11)+6\right)}{a_1^3} c_{{\emptyset}},\\
c_{\{2,1\}} & = \frac{{N} (\dvar +\beta  ({N}-1)+1) (\dvar +2 \beta  ({N}-1)+2) ({N} (\dvar +\beta  ({N}-1)+1)+2)}{a_1^3} c_{{\emptyset}},\\
c_{\{1,1,1\}} & = \frac{{N} (\dvar +\beta  ({N}-1)+1) ({N} (\dvar +\beta  ({N}-1)+1)+1) ({N} (\dvar +\beta  ({N}-1)+1)+2)}{a_1^3} c_{{\emptyset}},\\
c_{\{2\}} & = \frac{{N} (\dvar +\beta  ({N}-1)+1) (\dvar +2 \beta  ({N}-1)+2)}{a_1^2} c_{{\emptyset}},\\
c_{\{1,1\}} & = \frac{{N} (\dvar +\beta  ({N}-1)+1) ({N} (\dvar +\beta  ({N}-1)+1)+1)}{a_1^2} c_{{\emptyset}},\\
c_{\{1\}} & = \frac{{N} (\dvar +\beta  ({N}-1)+1)}{a_1} c_{{\emptyset}}~.
\end{aligned}
\end{equation}
Observe that all correlators of degree higher than 1 are proportional to $c_{\{1\}}$, which is a consequence of the fact that $\deg(W_{-1})=1$ and that there is only 1 partition in degree 1.
\\
\\
Our solution of the $\p=1$ model is slightly more general than the one in \cite{Itoyama:2017xid,Mironov:2017och} as we allow for a determinant deformation of parameter $\dvar$. For the special case $\dvar=0$ for the correlators above, we recover the formulas of \cite{Itoyama:2017xid}.

\subsubsection*{Averages of characters}

Another remarkable property of this model is that of \textit{super-integrability} \cite{Mironov:2017och,Mironov:2018ekq}, meaning that there are some observables whose expectation values satisfy a particularly nice formula. Namely, one observes that expectation values of characters can be expressed as simple combinations of the same characters evaluated at some specific ``points'' (see \cite{Morozov:2018eiq} for the original observation of this fact).
\\
\\
In the case of the $\beta$-deformed model, the natural characters to consider are the 1-parameter family of symmetric polynomials called Jack polynomials $\jack_\rho(p_k)$.
Using the solution we derived in \eqref{eq:WrepresentationP1}, one can explicitly check that
\begin{equation}
\label{eq:jackNf1}
 \lbracket\jack_\rho (p_k) \rbracket^{\p=1} =
 \frac{\jack_\rho(p_k=\beta^{-1}(\dvar+\beta(N-1)+1))}
 {\jack_\rho(p_k=\beta^{-1}{a_1}\delta_{k,1})}
 \jack_\rho(p_k=N) \, c_{\emptyset}~.
\end{equation}
Finally, in the limit $\beta=1$, the Jack polynomials degenerate to Schur polynomials (characters of the un-deformed model) whose averages satisfy the analogous relation
\begin{equation}
\label{eq:schurNf1}
\left.  \lbracket\schur_\rho (p_k) \rbracket^{\p=1} \right|_{\beta=1} =
 \frac{\schur_\rho(p_k=N+\dvar)}
 {\schur_\rho(p_k=a_1\delta_{k,1})}
 \schur_\rho(p_k=N) \, c_{\emptyset}~.
\end{equation}
While we are not aware of an analytical proof of these relations, we have been able to check that they hold for all partitions of degree 9 and lower.

\subsubsection{$\p = 2$}

The case of $\p = 2$ is a generalization of the familiar Hermitean Gaussian matrix model, with generating function given by
\begin{equation}
\label{eq:p2classicalgenfun}
 \Z^{\p = 2}(\tau) =
 \int_{\mathbb{R}^N}\prod_{i =
  1}^N \mathd \lambda_i \prod_{1 \leq i\neq j \leq N}(\lambda_i-\lambda_j)^{\beta}
 \mathe^{-a_1 \sum_{i=1}^N \lambda_i - \frac{1}{2}a_2 \sum_{i=1}^N \lambda_i^2
 + \sum_{s=1}^{\infty}\tau_s \sum_{i=1}^N \lambda_i^s}~.
\end{equation}
As we here require the $n=-1$ constraint in order to solve the model, we let $\dvar=0$ and thus we can re-sum all the constraints starting from $n=-1$. In order to obtain an operator $\constr$ whose diagonal is proportional to the dilatation operator $D$, we need to shift the weight of the re-summation as
\begin{equation}
\label{eq:p2classicalconstr}
 \constr = \sum_{n=-1}^\infty (n+2)\tau_{n+2}\constr_{n} = a_2 D - (W_{-2} - a_1 L_{-1})
\end{equation}
where
\begin{equation}
\begin{split}
 W_{-2} = & \beta \sum_{n,m=1}^\infty (n+m+2) \tau_{n+m+2} \partial_n\partial_m
 + (1-\beta)\sum^\infty_{n=1} (n+1)(n+2)\tau_{n+2}\partial_{n} + \\
 & + \sum_{n,m = 1}^\infty n m \tau_n \tau_m \partial_{n+m-2}
 + 2\beta N \sum_{n=1}^\infty (n+2)\tau_{n+2} \partial_n
 + (\beta N^2+(1-\beta)N) 2\tau_2 + \tau_1^2 N
\end{split}
\end{equation}
is an operator of degree 2, while $L_{-1}$ is defined as in \eqref{eq:virasoro_generators}. As in the previous case $D$ is the dilatation operator and it is of degree 0. An argument completely analogous to the one for $\p=1$ leads to the conclusion that $\constr$ is a triangular operator with a 1-dimensional kernel, and therefore that the solution to the equation $\constr\Z^{\p = 2}(\tau)=0$ is unique up to normalization.
\\
\\
In order to give a $W$-algebra representation of the generating function we first consider the simpler case of $a_1=0$, then we re-introduce the parameter $a_1$ by shifting $\tau_1\mapsto\tau_1-a_1$. For $a_1=0$ we have the Gaussian case originally solved in \cite{Morozov:2009xk} for which one can write
\begin{equation}
 \left. \Z^{\p = 2}(\tau) \right|_{a_1=0} = \exp\left(\frac{1}{2a_2}W_{-2}\right)\cdot  c_{\emptyset}~.
\end{equation}
Then we use the fact that
\begin{equation}
 [L_{-1},W_{-2}]=0
\end{equation}
together with the Virasoro constraint for $n=-1$ and $a_1=0$,
\begin{equation}
 a_2\partial_1 \left. \Z^{\p = 2}(\tau) \right|_{a_1=0} = L_{-1} \left. \Z^{\p = 2}(\tau) \right|_{a_1=0}
\end{equation}
to write the full solution as
\begin{equation}
\label{eq:WrepresentationP2}
\begin{aligned}
 \Z^{\p = 2}(\tau) & = \exp\left(-a_1\partial_1\right) \left[ \left. \Z^{\p = 2}(\tau) \right|_{a_1=0} \right] \\
 & = \exp\left(-\frac{a_1}{a_2}L_{-1}\right) \left[\left. \Z^{\p = 2}(\tau) \right|_{a_1=0} \right] \\
 & = \exp\left(\frac{1}{2a_2}W_{-2}-\frac{a_1}{a_2}L_{-1}\right)\cdot c_{\emptyset} ~.
\end{aligned}
\end{equation}
An explicit solution up to degree 3 is given by the correlators
\begin{equation}\label{eq:classical_correlators_p2}
\begin{aligned}
 c_{\{3\}} & = -\frac{a_1 {N} \left(3 a_2 ({\beta} ({N}-1)+1)+a_1^2\right)}{a_2^3} c_{\emptyset}~, \\
 c_{\{2,1\}} & = -\frac{a_1 {N} \left(a_2 \left({\beta} {N}^2-{\beta} {N}+{N}+2\right)+a_1^2 {N}\right)}{a_2^3} c_{\emptyset} ~,\\
 c_{\{1,1,1\}} & = -\frac{a_1 {N}^2 \left(a_1^2 {N}+3 a_2\right)}{a_2^3} c_{\emptyset} ~,\\
 c_{\{2\}} & = \frac{{N} \left(a_2 ({\beta} ({N}-1)+1)+a_1^2\right)}{a_2^2} c_{\emptyset} ~,\\
 c_{\{1,1\}} & = \frac{{N} \left(a_1^2 {N}+a_2\right)}{a_2^2} c_{\emptyset} ~,\\
 c_{\{1\}} & = -\frac{a_1 {N} }{a_2} c_{\emptyset}~.
\end{aligned}
\end{equation}

\subsubsection*{Averages of characters}

As observed in \cite{Mironov:2017och}, this model also satisfies the super-integrability property of characters. Using the solution derived in the previous section one can check that the following relation holds
\begin{equation}
\label{eq:jackNf2}
 \lbracket\jack_\rho (p_k) \rbracket^{\p=2} =
 \frac{\jack_\rho\left(p_k=(-1)^k\beta^{-1}({a_1\delta_{k,1}+a_2\delta_{k,2}})\right)}
 {\jack_\rho(p_k=\beta^{-1}{a_2}\delta_{k,1})}
 \jack_\rho(p_k=N) \, c_{\emptyset}~.
\end{equation}
Similarly, in the Schur limit where $\beta=1$ we have
\begin{equation}
\label{eq:schurNf2}
\left. \lbracket\schur_\rho (p_k) \rbracket^{\p=2} \right|_{\beta=1} =
 \frac{\schur_\rho\left(p_k=(-1)^k({a_1\delta_{k,1}+a_2\delta_{k,2}})\right)}
 {\schur_\rho(p_k=a_2\delta_{k,1})}
 \schur_\rho(p_k=N) \, c_{\emptyset}~,
\end{equation}
which is also consistent with the result of \cite{Morozov:2018eiq} (as given in \eqref{eq:schurMorozov}) when $a_1=0$. These relations have been checked for all partitions up to degree 9.

\subsubsection{Comments on $\p \geq 3$}
\label{sec:p>3}

Now consider higher values of $\p$ in \eqref{eq:beta_hermitean_mm_potential}.
By re-summing all constraints in \eqref{eq:virasoro_constraints} for $n\geq-1$ with weight $(n+\p)\tau_{n+\p}$ we obtain the equation
\begin{equation}
\label{eq:p>3constr}
 a_\p\underbrace{\left(D-\sum_{k=1}^{\p-2}k\tau_k\partial_k\right)}_{\text{diagonal}} \Z^{\p}(\tau)
 = \underbrace{\left(W_{-\p}-\sum_{k=1}^{\p-1}a_{\p-k} K_{-k}\right)}_{\text{off-diagonal}}\Z^{\p}(\tau)
\end{equation}
where the r.h.s. is a sum of shifted cut-and-join operators
\begin{equation}
\label{eq:Wp>3}
\begin{aligned}
 W_{-\p} = & (\p-1)\tau_1\tau_{\p-1} N
 + (\beta N^2+(1-\beta)N) \p\tau_\p
 + \sum_{n=1}^\infty\sum_{m=\p-1}^\infty n m \tau_n\tau_m\partial_{n+m-\p} \\
 & + \sum_{n=1}^\infty\left[2\beta N+(1-\beta)(n+1)\right](n+\p)\tau_{n+\p}\partial_n
 + \beta\sum_{n=1}^\infty\sum_{m=1}^\infty (n+m+\p)\tau_{n+m+\p}\partial_n\partial_m ~,
\end{aligned}
\end{equation}
and
\begin{equation}
 K_{-k} := \sum_{n=\p-1}^\infty n \tau_n \partial_{n-k}
 + \delta_{k,\p-1}(\p-1)\tau_{\p-1}N~,
\end{equation}
with $\deg W_{-\p} = \p$ and $\deg K_{-k} = k$.
The l.h.s. of \eqref{eq:p>3constr} is of degree zero, therefore it corresponds to the diagonal part of the triangular operator $\constr=\sum_{n\geq-1}(n+\p)\tau_{n+\p}\constr_n$, while $W_{-\p}$ and $K_{-k}$ are of positive degree and therefore they represents the off-diagonal part of $\constr$.
\\
\\
For $\p\geq3$ however, we immediately notice that the kernel of $\constr$ is of dimension greater than 1. The kernel of the operator $D-\sum_{k=1}^{\p-2}k\tau_k\partial_k$ is in fact infinite dimensional and corresponds to the span of all monomials which do not contain times $\tau_k$ for $k>\p-2$.
For example, if $\p=3$ all monomials of the form $\tau_1^\ell$ for all positive integer powers $\ell$ are annihilated by the diagonal part of $\constr$. This means that equation \eqref{eq:p>3constr} does not provide a recursion relation expressing the corresponding correlator $c_{\{1,\dots,1\}}$ as a linear combination of correlators of lower degree.
Consequently one should consider these coefficients as additional background data that needs to be specified independently in order to fully determine the generating function. As remarked in \cite{Cordova:2016jlu}, for finite values of $N$ one can always find additional relations between such correlators because only at most $N$ of those can be linearly independent for a matrix of finite size. Therefore one can reduce the indeterminacy of the system of equations coming from the Virasoro constraints to a finite amount of information. Nevertheless, one cannot write a full solution for the generating function either in terms of correlators or in $W$-algebra representation. 
\\
\\
We now present a formal way to repackage all the information that can be obtained from the recursion (for earlier attempts see \cite{Alexandrov:2003pj,Alexandrov:2004ed,Alexandrov:2004ud}). From the integral representation of the generating function we can derive the additional identities
\begin{equation}
\label{eq:wardp3}
 \left(\frac{\partial}{\partial\tau_k}+k\frac{\partial}{\partial a_k}\right)\Z^{\p}(\tau;a)=0~,
 \quad\quad\quad k=1,\dots,\p
\end{equation}
where we have also explicitly written the dependence of the generating function on the coupling constants $a_k$. If we substitute \eqref{eq:wardp3} in the l.h.s. of \eqref{eq:p>3constr} we can rewrite the term $-\sum_{k=1}^{\p-2}k\tau_k\partial_k$ as the operator
\begin{equation}
 \sum_{k=1}^{\p-2}k^2\tau_k\frac{\partial}{\partial a_k}~,
\end{equation}
which is now no longer of zero degree in the times (in fact, since $\partial/\partial a_k$ has degree zero, every term in the sum has the same degree as $\tau_k$) which means that it is not a diagonal operator.
If we write $\widetilde{W}$ for the off-diagonal part of $\constr$,
\begin{equation}
 \widetilde{W} := \frac{W_{-\p}}{a_\p}-\sum_{k=1}^{\p-1}\frac{a_{\p-k}}{a_\p} K_{-k}
 - \sum_{k=1}^{\p-2}k^2\tau_k\frac{\partial}{\partial a_k}~,
\end{equation}
we have that \eqref{eq:p>3constr} becomes
\begin{equation}
 D\Z^{\p}(\tau;a) = \widetilde{W}\Z^{\p}(\tau;a)~.
\end{equation}
This constraint is still triangular (with respect to a basis of $\mathbb{C}[[\tau_1,\tau_2,\dots]]$) but now its diagonal component is the operator $D$, which we know has 1-dimensional kernel and in particular it is invertible over the complement of its kernel.
A formal solution can now be obtained by splitting the generating function as
\begin{equation}
 \Z^{\p}(\tau;a) = c_{\emptyset}(a) + \Z^{\p}_{\perp}(\tau;a)~,
\end{equation}
where $c_{\emptyset}(a)\equiv\Z^{\p}(0;a)$ is the component of $\Z^{\p}(\tau;a)$ which sits in the kernel of $D$, while $\Z^{\p}_{\perp}(\tau;a)$ is the component which sits in the complement of $\ker D$ (i.e. $\Z^{\p}_{\perp}(0;a)=0$). Then we can write
\begin{equation}
 \left(D-\widetilde{W}\right)\Z^{\p}_{\perp}(\tau;a) = \widetilde{W} c_{\emptyset}(a)~,
\end{equation}
and observing that $D$ and $(D-\widetilde{W})$ are invertible operators when restricted to the image of $\widetilde{W}$ (which is contained in the complement of $\ker D$), we obtain
\begin{equation}
\label{eq:formalsolutionp}
\begin{aligned}
 \Z^{\p}(\tau;a) & = \left(1+(D-\widetilde{W})^{-1} \widetilde{W} \right) c_{\emptyset}(a) \\
 & = \left(1+(1-D^{-1}\widetilde{W})^{-1} D^{-1}\widetilde{W} \right) c_{\emptyset}(a) \\
 & = \sum_{n=0}^\infty (D^{-1}\widetilde{W})^{n} c_{\emptyset}(a)~.
\end{aligned}
\end{equation}
This formal expression for the generating function automatically implements the additional constraints \eqref{eq:wardp3} but only for $1\leq k\leq \p-2$, precisely because in the operator $\widetilde{W}$ the derivatives with respect to $a_{\p-1}$ and $a_{\p}$ do not appear.
This implies that our solution \eqref{eq:formalsolutionp} in general does not satisfy \eqref{eq:wardp3} if $k=\p-1$ or $k=\p$. Without loss of generality then we can assume $a_{\p}=1$ and $a_{\p-1}=0$ and repeat the argument that leads to \eqref{eq:formalsolutionp}. The final answer now is totally unambiguous and only depends on an appropriate choice of the correlation function
\begin{equation}
\label{eq:emptycorr}
 c_{\emptyset}(a_1,\dots,a_{\p-2}) \equiv
 \int_{\mathbb{R}_{>0}^N}\prod_{i=1}^N \mathd\lambda_i\,
 \Delta(\lambda)^{2\beta} \exp\left(
 - a_{1}\sum_{i=1}^N\lambda_{i}
 - \dots
 - \frac{a_{\p-2}}{\p-2}\sum_{i=1}^N \lambda_{i}^{\p-2}
 - \frac{1}{\p}\sum_{i=1}^N \lambda_{i}^{\p}
 \right)~.
\end{equation}
Because $\widetilde{W}$ contains derivatives in the variables $a_k$, the recursion relations are no longer polynomial in the correlators $c_{\rho}(a)$. In fact we have that $c_{\emptyset}(a)$ acts as a generating function for all the correlators that the recursion could not fix and the formal solution \eqref{eq:formalsolutionp} expresses them as $a$-derivatives of $c_{\emptyset}(a)$.
\\
\\
For example, if $\p=3$ we can write the solution up to degree 4 as
\begin{equation}
\begin{aligned}
c_{\{4\}} & = \left(a_1^2 {N}-2 (\beta ({N}-1)+1) \frac{\partial}{\partial a_1}\right)c_{\emptyset}(a_1), \\
c_{\{3,1\}} & = -\left(\left(\beta {N}^2-\beta {N}+{N}+1\right) \frac{\partial}{\partial a_1}
 +a_1 \left(\frac{\partial}{\partial a_1}\right)^2 \right)c_{\emptyset}(a_1), \\
c_{\{2,2\}} & = \left(a_1^2 {N}^2 - 2 \frac{\partial}{\partial a_1}\right)c_{\emptyset}(a_1), \\
c_{\{2,1,1\}} & = -{N}\left(2\frac{\partial}{\partial a_1}+a_1 \left(\frac{\partial}{\partial a_1}\right)^2\right)c_{\emptyset}(a_1), \\
c_{\{1,1,1,1\}} & = \left(-\frac{\partial}{\partial a_1}\right)^4 c_{\emptyset}(a_1), \\
c_{\{3\}} & = \left((\beta ({N}-1) {N}+{N}) +a_1 \frac{\partial}{\partial a_1}\right)c_{\emptyset}(a_1), \\
c_{\{2,1\}} & = {N} \left(1+a_1 \frac{\partial}{\partial a_1}\right)c_{\emptyset}(a_1), \\
c_{\{1,1,1\}} & = \left(-\frac{\partial}{\partial a_1}\right)^3 c_{\emptyset}(a_1), \\
c_{\{2\}} & = -a_1{N} c_{\emptyset}(a_1), \\
c_{\{1,1\}} & = \left(-\frac{\partial}{\partial a_1}\right)^2 c_{\emptyset}(a_1), \\
c_{\{1\}} & = -\frac{\partial}{\partial a_1} c_{\emptyset}(a_1) .
\end{aligned}
\end{equation}
It is curious to notice that while the integral representation of \eqref{eq:emptycorr} is natural from the point of view of the definition of the matrix model, the solution of the Virasoro constraints does make sense also for an arbitrary function $c_{\emptyset}(a_1,\dots,a_{\p-2})$ which does not necessarily admit an integral representation of that form.

\section{Quantum models}
\label{sec:quantum_models}

We now shift our attention to the $q$-deformation of the classical models presented in the previous section. These will correspond to families of deformations depending on 1 or more parameters (typically $q,t$ and in some cases $r$) which in the limit of those parameters going to 1 reduce the familiar examples already discussed. We refer to the $q$-deformed models as the \textit{quantum} version of the Hermitean matrix model and to the degeneration
limit $q\to1$ as their \textit{semi-classical} approximation.
\\
\\
One more motivation for the name ``quantum'' is that we are able to identify such matrix models as the localized partition functions of certain supersymmetric quantum field theories in 3 dimensions. More explicitly, these correspond to theories with 4 supercharges placed on backgrounds of the form $\halfindex$ or $S^3_b$. As explained below, in some specific sense one can regard the $\halfindex$ partition function as a half of the partition function on $S^3_b$.
\\
\\
The $q$-deformation of the classical Virasoro constraints is a system of finite difference equations obtained by acting with operators satisfying a $q$-analogue of the Virasoro algebra. In the following, we mimic the derivation and solution of the Virasoro constraints in the $q$-case while simultaneously providing a detailed matching of the parameters between the quantum and the semi-classical case. Schematically, we have the identifications of parameters as in Table~\ref{tab:scheme}.
\begin{table}[ht]
\caption{Matching of parameters between quantum and classical models. {While the integer parameters $N_f$ and $\p$ can be straightforwardly identified, for the other parameters the identification is slightly less obvious. The $\beta$-deformation is obtained by identifying the adjoint mass as $t=q^\beta$ while $\dvar$ can be related to the effective FI parameter through the parameter $r	$. Similarly, the coupling constants $a_k$ are given by non-trivial functions of the masses $u_k$.}}
\label{tab:scheme}
\renewcommand\arraystretch{1.5}
\noindent
\begin{center}
\begin{tabular}{|cc|cc|}
 \hline
 \multicolumn{2}{|c|}{Quantum model} & \multicolumn{2}{|c|}{Classical model} \\
 \hline
 (adjoint mass)& $t$ & $\beta$ & ($\beta$-deformation)\\
 (number of flavors)& $N_f$ & $\p$ & (degree of potential)\\
 (fundamental masses)& $u_k$ & $a_k$ & (coupling constants)\\
 (balancing parameter)& $r$ & $\dvar$ & (determinant insertion)\\
 \hline
\end{tabular}
\end{center}
\end{table}

\subsection{Definitions}
We will now provide the details of the $q$-models which we intend to study, namely certain supersymmetric gauge theories in three dimensions.

\subsubsection{$\halfindex$}

We consider the partition function of a 3d $\mathcal{N}=2$ supersymmetric theory with gauge group $U(N)$ on the 3-manifold $\halfindex$. More precisely, the geometry is that of a $D^2$ fibration over $S^1$ such that the $D^2$ fiber is rotated of a parameter $q$ when going around the base.
The $U(1)$ holonomy $q$ is identified with the quantum deformation parameter of the resulting matrix model. The partition function of an $\mathcal{N}=2$ theory on this geometry is sometimes referred to as the \textit{half-index} \cite{Dimofte:2011py} or \textit{holomorphic block} \cite{Beem:2012mb}.
\\
\\
Besides the $\mathcal{N}=2$ vector multiplet, we consider also an adjoint chiral multiplet of mass $t$ and $N_f$ fundamental anti-chiral fields of masses $u_k$. Moreover, we also turn on a Fayet-Iliopoulos (FI) parameter $\kappa_1 \in \mathbb{C}$.
The partition function can be computed via supersymmetric localization \cite{Beem:2012mb,Yoshida:2014ssa} with the result
\begin{equation}
\label{eq:D2xS1partition_function}
 \Z_{\halfindex} = \oint_\C \prod_{i=1}^N \frac{\mathd\lambda_i}{\lambda_i}
 \Z_{\halfindex}^{\mathrm{cl}}(\lambda ) \Z_{\halfindex}^{\mathrm{1-loop}}(\lambda )~,
\end{equation}
where the contour $\C$ is a middle dimensional cycle in $(\mathbb{C}^\times)^N$ defined by taking the product of $N$ copies of the unit circle. Observe that there might be analytical issues with this naive choice of contour when the parameters are non-generic (see \cite{Beem:2012mb}).
{However we will avoid discussing such difficulties and assume that an appropriate contour exists by defining the generating function as an analytically well-defined solution to a set of partial differential equations obtained via algebraic manipulations of the integrand. More specifically, these equations will be the $q$-Virasoro constraints discussed in Section~\ref{sec:q_vir_constraints}. In the generic case the two approaches are equivalent.}
\\
\\
The integrand of the partition function is defined as the product of the classical contribution
\begin{equation}
 \Z_{\halfindex}^{\mathrm{cl}}(\lambda)= \prod_{i=1}^N \lambda_i^{\kappa_1}
\end{equation}
and the product of 1-loop determinants
\begin{equation}
 \Z_{\halfindex}^{\mathrm{1-loop}}(\lambda) = \prod_{1 \leq k \neq l \leq N}
 \frac{(\lambda_k/\lambda_l;q)_\infty}{(t\lambda_k/\lambda_l;q)_\infty}
 \prod_{j=1}^N \prod_{k=1}^{N_f} (q\lambda_j u_k;q)_\infty~,
\end{equation}
coming from the contributions of the vector, adjoint chiral and fundamental anti-chiral multiplets.
Here $(x;q)_\infty$ is the $q$-Pochhammer symbol defined in \eqref{eq:qpochhammer}. 
\\
\\
The partition function in \eqref{eq:D2xS1partition_function} can then be interpreted as the matrix model where the measure
\begin{equation}
 \Delta_{q,t}(\lambda) = \prod_{1 \leq k \neq l \leq N} \frac{(\lambda_k/\lambda_l;q)_\infty}
 {(t\lambda_k/\lambda_l;q)_\infty} = \prod_{1 \leq k \neq l \leq N} \prod_{n = 0}^{\infty } \frac{1-\lambda_k / \lambda_l q^n}{1-t \lambda_k / \lambda_l q^n}
\end{equation}
can be seen as the $q$-deformed Vandermonde determinant, while the remaining contributions can be interpreted as $q$-deformations of the classical potential $V(\lambda_i)$. The contribution $\Z_{\halfindex}^{\mathrm{cl}}(\lambda)$ for instance has the form of a determinant insertion, however the actual exponential of the determinant will receive corrections from the measure $\Delta_{q,t}(\lambda)$.
In order to make the connection to the potential in \eqref{eq:potential} we observe that the 1-loop determinant of the fundamental multiplets can be formally rewritten using the identity \eqref{eq:qpochhammeridentity} as
\begin{equation}
\label{eq:shiftoftimesD2S1}
 \prod_{j=1}^N\prod_{k=1}^{N_f}(q\lambda_j u_k;q)_\infty =
 \exp\left(-\sum_{s=1}^\infty\frac{p_{s}(u)}{s(q^{-s}-1)}\sum_{i=1}^N\lambda_i^s\right)~,
\end{equation}
where $p_s(u)$ are the power-sum variables for the masses $u_k$,
\begin{equation}
\label{eq:mass_powersum}
 p_s(u) := \sum_{k=1}^{N_f} u_k^s ~. 
\end{equation}
Once we define the generating function
\begin{equation}
\label{eq:generating_function}
 \Z_{\halfindex}^{N_f}(\tau) = \oint_\C \prod_{i=1}^N \frac{\mathd\lambda_i}{\lambda_i}
 \Z_{\halfindex}^{\mathrm{cl}}(\lambda) \Z_{\halfindex}^{\mathrm{1-loop}}(\lambda )\,\mathe^{\sum_{s=1}^\infty \tau_s \sum_{i=1}^N \lambda_i^s}~,
\end{equation}
by introducing the standard coupling to times $\{\tau_s\}$, we can see that \eqref{eq:shiftoftimesD2S1} is a ``potential'' of the form which can be obtained by the shift of times
\begin{equation}
\label{eq:shiftoftimesD2S1other}
 \tau_s \mapsto \tau_s - \frac{p_{s}(u)}{s(q^{-s}-1)}~.
\end{equation}
Notice however that here all of the times must be shifted, as opposed to the classical case where only a finite number (corresponding to the integer $\p$) had a non-trivial shift.
\\
\\
For a generic (polynomial) operator $\mathcal{O}=\mathcal{O}(\lambda)$ we define its expectation value using the notation
\begin{equation}
\label{eq:expectationvalue}
 \lbracket \mathcal{O} \rbracket_{\tau}^{N_f} =
 \oint_\C \prod_{i=1}^N \frac{\mathd\lambda_i}{\lambda_i}\mathcal{O}(\lambda)
 \Z_{\halfindex}^{\mathrm{cl}}(\lambda)\Z_{\halfindex}^{\mathrm{1-loop}}(\lambda)\,
 \mathe^{\sum_{s=1}^\infty\tau_s\sum_{i=1}^N \lambda_i^s}~,
\end{equation}
where $\lbracket\mathcal{O}\rbracket^{N_f}$ is obtained by setting all the times to zero in the previous formula. Moreover, we assume that $\Z_{\halfindex}^{N_f}(\tau)$ admits a formal power series expansion in times, whose coefficients are the correlators $c_\rho$ of the theory.
\\
\\
We pause here to explain the physical meaning of the generating function $\Z_{\halfindex}^{N_f}(\tau)$. From the point of view of the gauge theory on $\halfindex$, one is interested in computing expectation values of gauge invariant quantities such as the Wilson loops. These correspond to characters of $U(N)$ evaluated on the holonomy of the gauge connection around some BPS closed curve. In the case of the background at hand, there is one BPS loop corresponding to the zero section of the $D^2$ bundle over $S^1$, and it is invariant under the $U(1)$ action on the fibers. Since characters of the unitary group are given by the Schur polynomials, one can write any Wilson loop expectation value as the average of some Schur polynomial written on the basis of power-sum variables $p_s$,
\begin{equation}
\label{eq:wilsonloop_d2s1}
 \mathrm{WL}_\rho
 = \langle\schur_\rho (p_k=\sum_i\lambda_i^k) \rangle^{N_f}~,
\end{equation}
where $\rho$ is the integer partition labeling the highest weight of the representation (see Appendix~\ref{sec:characters} for a short review of symmetric functions and Schur polynomials).

\subsubsection{$S_b^3$}

A different but intimately related model is that of an $\mathcal{N}=2$ YM-CS theory on the squashed 3-sphere $S^3_b$. We consider $U(N)$ gauge group and the same matter content as before. The 3d geometry is that defined by the equation
\begin{equation}
 \omega_1 |z_1|^2 + \omega_2 |z_2|^2 = 1, \quad\quad\quad z_1,z_2\in\mathbb{C}
\end{equation}
where $\omega_1,\omega_2\in\mathbb{R}$ are the squashing parameters. The dependence of the partition function on the squashing is often indicated via a real parameter $b$ such that $b^2 = \omega_2/\omega_1$ \cite{Kapustin:2009kz, Hama:2010av, Imamura:2011wg, Alday:2013lba}. {We remark that, while geometrically it is natural to take the squashing parameters to be real valued, most of the formulas that we write in this paper are well-defined for arbitrary complex values. From now on, unless explicitly specified, we will assume $\omega_1,\omega_2\in\mathbb{C}$.}
\\
\\
Topologically, we can think of the 3-sphere as the gluing of two solid tori, i.e. two copies of $D^2\times S^1$ whose boundaries are identified via a modular transformation which acts by exchanging the two fundamental cycles of $T^2$. Each half of the sphere can then be though to define a copy of the theory on $D^2\times_{q_\alpha}S^1$ where now each copy has its own modular parameter $q_\alpha$, $\alpha=1,2$, which we can express through the squashing parameters of the sphere as
\begin{equation}
 q_1 = \mathe^{2\pi\mathi\frac{\omega_2}{\omega_1}},
 \quad\quad\quad
 q_2 = \mathe^{2\pi\mathi\frac{\omega_1}{\omega_2}}~.
\end{equation}
This simple geometric picture eventually leads to the very non-trivial property of factorization of the $S^3_b$ partition function into a product of holomorphic blocks \cite{Pasquetti:2011fj}. Here we are interested in yet another consequence of this factorization, namely the fact that the 3-sphere partition function satisfies two independent sets of $q$-Virasoro constraints as shown in \cite{Nedelin:2016gwu}, hence the name \textit{modular double}.
\\
\\
As in the previous model, the theory on $S^3_b$ has an $\mathcal{N}=2$ vector, an adjoint chiral of mass $M_\mathrm{a}$ and $N_f$ anti-chiral fundamental fields of masses $m_k$. Again, we allow for a non-zero FI parameter $\kappa_1$ however in this case we are also forced to introduce a non-vanishing (bare) Chern-Simons (CS) level $\kappa_2$. The reason for this is not physical in nature but rather it arises as a technical requirement necessary for having $q$-Virasoro constraints which can be written as PDEs in the time variables. In the case $N_f=2$ this was first shown in \cite{Cassia:2019sjk} where a unit CS level had to be introduced.
Here we generalize that condition to arbitrary number of flavors $N_f\geq1$ by imposing that
\begin{equation}
\label{eq:effectiveCS}
 N_f = 2 \kappa_2~,
\end{equation}
or equivalently, that the effective\footnote{In the presence of matter fields, the CS level receives quantum corrections, so that one can define an effective CS level $\kappa_2^\mathrm{eff}\in\mathbb{Z}$, which then has to satisfy a quantization condition for the theory to be free of anomalies. In particular, this implies that the bare CS level $\kappa_2$ can be taken to be an half-integer number as long as we have an appropriate number of matter fields.} CS level $\kappa_2^\mathrm{eff}:=\kappa_2-N_f/2$ be vanishing. Observe that this condition is compatible with the cancellation of all perturbative anomalies even when the bare CS level is half-integral (i.e. $N_f$ is odd). We remark also that a non-zero effective CS level would correspond, from the point of view of the classical matrix model, to a potential term of the form $-\log^2(\lambda)$, which would similarly spoil the derivation of the usual Virasoro constraints.
\\
\\
The partition function can be computed by means of supersymmetric localization techniques \cite{Imamura:2011wg,Hama:2011ea} and the result is given by
\begin{equation}
\label{eq:partitionfunction}
 \Z_{S_b^3} = \int_{(\mathi\mathbb{R})^N} \prod_{i=1}^N \mathd X_i
 \,\Z_{S_b^3}^{\mathrm{cl}}(X)\, \Z_{S_b^3}^{\mathrm{1-loop}}(X)~,
\end{equation}
where $X_i\in\mathi\mathbb{R}$ are Coulomb branch variables, $\Z_{S_b^3}^{\mathrm{cl}}(X)$ is the classical contribution
\begin{equation}
 \Z_{S_b^3}^{\mathrm{cl}}(X) =  \prod_{i=1}^N
 \exp\left(-\frac{\pi\mathi\kappa_2}{\omega_1 \omega_2}X_i^2
 +\frac{2 \pi \mathi \kappa_1}{\omega_1\omega_2} X_i \right)
\end{equation}
and $\Z_{S_b^3}^{\mathrm{1-loop}}(X)$ is the product of 1-loop determinants
\begin{equation}
 \Z_{S_b^3}^{\mathrm{1-loop}}(X) =
 \prod_{ 1 \leq k\neq j \leq N}\frac{S_2(X_k-X_j|\underline{\omega})}{S_2(X_k-X_j+M_{\mathrm{a}}|\underline{\omega})}
 \prod_{k=1}^{N_f} \prod_{i = 1}^N
 S_2 \left(-X_i-m_k|\underline{\omega} \right)^{-1}
\end{equation}
which is written in terms of the double sine function $S_2(z|\underline{\omega})$ defined in \eqref{eq:doubleSine}. Analytical issues related to the convergence of the integral and its dependence on the physical parameters are addressed in Appendix~\ref{sec:asymptotic_analysis}.
\\
\\
{It is worth mentioning that for $N_f=2,3$ there are some known dualities for the 3d partition function on $S^3_b$. More specifically, in \cite{Amariti:2018wht} it was observed that the $U(N)_1$ theory with 2 anti-chirals is dual to $U(N)_1$ with 1 chiral and 1 anti-chiral as well as to $USp(2N)_4$ with 2 flavors \cite[Theorem 5.6.19]{vdBult:2007}. The $U(N)_{3/2}$ theory with 3 anti-chirals is instead dual to $U(N)_0$ with 2 anti-chirals \cite[Theorem 5.6.20]{vdBult:2007}. Unfortunately, we are not aware of other dualities of this type for arbitrary number of flavors $N_f$. It would be interesting to understand the meaning of these dualities from the matrix model point of view, however we will not address this question in the present paper.}
\\
\\
In order to simplify the notation we introduce the following set of exponentiated variables
\begin{equation}
\label{eq:exponentiated_variables}
\begin{aligned}
\begin{aligned}
 q_{\alpha} &:= \mathe^{\frac{2 \pi \mathi \omega}{\omega_{\alpha}}} \\
 t_{\alpha} &:= \mathe^{\frac{2 \pi \mathi  M_{\mathrm{a}}}{\omega_{\alpha}}} \\
\end{aligned}
&\quad\quad\quad
\begin{aligned}
 u_{k,\alpha} &:= \mathe^{\frac{2 \pi \mathi m_k}{\omega_\alpha}} \\
 \lambda_{i,\alpha} &:= \mathe^{\frac{2 \pi \mathi X_i}{\omega_{\alpha}}}~,
\end{aligned}
\end{aligned}
\end{equation}
with $\omega=\omega_1+\omega_2$. Introducing the complex parameter $\beta=M_\mathrm{a}/\omega$, we can also write $t_\alpha=q_\alpha^\beta$.
\\
\\
The partition function in \eqref{eq:partitionfunction} then defines a quantum deformation of an eigenvalue matrix model with measure
\begin{equation}
 \Delta_S (X) = \prod_{1 \leq k\neq j \leq N}\frac{S_2(X_k-X_j|\underline{\omega})}
 {S_2(X_k-X_j+M_{\mathrm{a}}|\underline{\omega})}
\end{equation}
and potential
\begin{equation}
 \prod_{k=1}^{N_f} \prod_{i = 1}^N
 S_2 \left(-X_i-m_k|\underline{\omega} \right)^{-1}
 \prod_{i=1}^N
 \exp\left(-\frac{\pi\mathi\kappa_2}{\omega_1 \omega_2}X_i^2
 +\frac{2 \pi \mathi \kappa_1}{\omega_1\omega_2} X_i \right)~.
\end{equation}
Observe that, for $\omega_{1,2}$ in generic positions, one can use the identity \eqref{eq:S2identity} and regard the double sine function as the product of two $q$-Pochhammer symbols with arguments $\lambda_{i,1}$ and $\lambda_{i,2}$, respectively.
This can be used to argue that the $S^3_b$ generating function satisfies two independent set of $q$-Virasoro constraints, however it can be shown (see \cite{Cassia:2019sjk}) that this latter property holds for any value of the squashing parameters, even when \eqref{eq:S2identity} does not apply.
\\
\\
Because of the presence of two separate (but not independent) sets of variables $\{\lambda_{i,\alpha}\}$, we can define a ``doubled'' generating function by coupling each set to its own copy of auxiliary time variables,
\begin{equation}
\label{eq:generatingfunction}
 \Z_{S_b^3}^{N_f}(\tau) = \int_{(\mathi\mathbb{R})^N} \prod_{i=1}^N \mathd X_i
 \,\Z_{S_b^3}^{\mathrm{cl}}(X)\, \Z_{S_b^3}^{\mathrm{1-loop}}(X)
 \prod_{\alpha=1,2} \exp \left( \sum_{s = 1}^{\infty} \tau_{s, \alpha}
 \sum_{i=1}^N \lambda_{i, \alpha}^s \right)
\end{equation}
where now $\{\tau_{s,\alpha}\}$ act as conjugate variables for the power-sum observables 
\begin{equation}
\label{eq:powersum}
 p_{s,\alpha} = \sum_{i=1}^N \lambda_{i, \alpha}^s~.
\end{equation}
As a formal power series in times, the generating function can be written as
\begin{equation}
\label{eq:formalseriesS3}
 \Z_{S_b^3}^{N_f}(\tau) = \sum_{\rho}\sum_{\rho'}
 \frac{1}{|\Aut(\rho )|}\frac{1}{|\Aut(\rho')|}
 c_{\rho;\rho'}
 \prod_{a\in\rho }\tau_{a,1}
 \prod_{b\in\rho'}\tau_{b,2}
\end{equation}
for $\rho,\rho'$ integer partitions of arbitrary sizes and
\begin{equation}
\label{eq:s3_correlators}
 c_{\rho;\rho'} = \left. \left[\prod_{a\in\rho}
 \frac{\partial}{\partial\tau_{a,1}}\prod_{b\in\rho'}
 \frac{\partial}{\partial\tau_{b,2}}\right] \Z_{S_b^3}^{N_f}(\tau)\right|_{\tau=0}
 = \lbracket \prod_{a\in\rho}p_{a,1}\prod_{b\in\rho'} p_{b,2} \rbracket~.
\end{equation}
are the correlation functions of products of the power-sums.
\\
\\
Expectation values of Wilson loops are defined equivalently to the case of $\halfindex$ given in \eqref{eq:wilsonloop_d2s1}, however they now carry a dependence on $\alpha$ due to the power-sum variables \eqref{eq:powersum} in the argument of the Schur polynomial. Physically, this corresponds to the fact that on $S^3_b$ there are two BPS cycles, one for each solid torus, and therefore the holonomy of the gauge connection must carry an index $\alpha$ distinguishing between the two.
\\
\\
It is now worth noting that we can formally obtain the generating function on $\halfindex$ \eqref{eq:generating_function} from the generating function on $S^3_b$ \eqref{eq:generatingfunction} by setting one set of times to zero, $\{ \tau_{s,2}=0 \}$ for instance,
\begin{equation}
\label{eq:S3andD2S1}
 \left. \Z^{N_f}_{S_b^3}(\tau)\right|_{\tau_{s,2}=0} \simeq \Z^{N_f}_{\halfindex}(\tau) ~.
\end{equation}
These generating functions are then equivalent in the sense that they satisfy the same set of constraints as will be shown below in \eqref{eq:qVirasoroD2S1} and \eqref{eq:qVirasoroS3}, however they do not have the same integral representation. Thus, whenever we encounter ambiguities in defining the contour $\mathcal{C}$ in the partition function of $\halfindex$ in \eqref{eq:D2xS1partition_function} we can resolve them by choosing any contour which is consistent with \eqref{eq:S3andD2S1}.
In other words, we can choose any contour such that the coefficients in the power series expansion of $\Z^{N_f}_{\halfindex}(\tau)$ are convergent (as integrals) as long as the resulting generating function satisfies the same set of PDEs as half of $\Z^{N_f}_{S_b^3}(\tau)$.

\subsection{$q$-Virasoro constraints}
\label{sec:q_vir_constraints}

Mirroring the procedure of the classical case, we will now derive constraint equations for the gauge theory on $\halfindex$ and then briefly summarize the corresponding procedure in the case of $S_b^3$. {The techniques used to derive the constraints are those developed in \cite{Lodin:2018lbz,Cassia:2019sjk} therefore we avoid presenting each step of the computation. The most important differences with respect to those cases are the presence of an arbitrary number of flavors $N_f$ and the introduction of the additional deformation parameter $r$.}

\subsubsection{$\halfindex$}

We now show that the generating function \eqref{eq:generating_function}
satisfies a set of first order $q$-difference equations which take the name of $q$-Virasoro constraints.
In order to derive these $q$-Virasoro constraints we introduce the finite difference operator $\newshift$ defined as
\begin{equation}
\label{eq:shiftoperator}
 \newshift f(\lambda ) = f( \dots, q ^{-1} \lambda_i , \dots ) 
\end{equation}
for $f$ a function of the gauge variables $\{\lambda_i\}$.
The constraints are obtained by substituting in \eqref{eq:totalvariation} the partial derivative $\partial/\partial\lambda_i$ with the difference operator\footnote{Observe that in the semi-classical limit $t\to1$ and $q\to1$, the function $G_i(\lambda;t)$ goes to 1 while \[\lim_{q\to1}\frac{(\newshift-1)}{\lambda_i(q^{-1}-1)}=\frac{\partial}{\partial\lambda_i}~.\]} $(\newshift-1)G_{i}(\lambda;t)$ where
\begin{equation}
 G_{i}(\lambda ;t) = \prod_{j\neq i}\frac{1-t\lambda_{i}/\lambda_{j}}{1-\lambda_{i}/\lambda_{j}} ~.
\end{equation}
The vanishing of the $q$-variation of the generating function can then be written as
\begin{equation}
\label{eq:finitediff}
  \oint_\C \prod_{i=1}^N \frac{\mathd\lambda_i}{\lambda_i}\sum_{i=1}^{N}
  \left[\lambda_{i}^n G_{i}(\lambda;t)\dots\right]
  = \oint_\C \prod_{i=1}^N \frac{\mathd\lambda_i}{\lambda_i}\sum_{i=1}^{N}\newshift
  \left[\lambda_{i}^n G_{i}(\lambda ;t)\dots\right]~,
\end{equation}
with ``$\dots$'' denoting the integrand.
\\
\\
For the sake of simplicity of the computation, we sum all such equations for the integer $n$ ranging over all of $\mathbb{Z}$ and multiply each component by the corresponding power $z^n$ of an auxiliary formal variable $z$. We finally end up with the single $q$-Virasoro constraint equation
\begin{equation}
\begin{split}
  \left(\mathrm{LHS}\right){:=} &
  \oint_\C \prod_{i=1}^N \frac{\mathd \lambda_i}{\lambda_i} \sum_{i=1}^{N}\left[ \sum_{n\in\mathbb{Z}} (z\lambda_{i})^n
  G_{i}(\lambda ;t)\right]\Z_{\halfindex}^{\mathrm{cl}}(\lambda ) \Z_{\halfindex}^{\mathrm{1-loop}}(\lambda)\mathe^{\sum_{s=1}^\infty \tau_s \sum_{i=1}^N \lambda_i^s}
  = \\
  & = \oint_\C \prod_{i=1}^N \frac{\mathd \lambda_i}{\lambda_i} \sum_{i=1}^{N}\newshift \left[
  \sum_{n\in\mathbb{Z}} (z\lambda_{i})^n
  G_{i}(\lambda ;t)\Z_{\halfindex}^{\mathrm{cl}}(\lambda ) \Z_{\halfindex}^{\mathrm{1-loop}}(\lambda)\mathe^{\sum_{s=1}^\infty \tau_s \sum_{i=1}^N \lambda_i^s} \right]
  {=:}\left(\mathrm{RHS}\right)
\end{split}
\end{equation}
and each independent constraint can be recovered by expanding in the given power of $z$.
\\
\\
The LHS and RHS can be computed independently. Starting with the RHS we have
\begin{equation}
\begin{aligned}
 (\mathrm{RHS}) =& 
q^{-\kappa_1} \exp\left( \sum_{s=1}^\infty z^{-s}(1-q^s)\left(\tau_{s}
 +\frac{p_{s}(u)}{s(1-q^{-s})}\right) \right)\times \\
 \times& \left\langle
 \frac{1}{1-t}
 \exp\left(\sum_{s=1}^\infty z^{s} \frac{(1-t^s)}{s q^s}
 \sum_{i=1}^N\lambda_{i}^{s} \right)
 -\frac{t^N}{1-t}\exp\left(\sum_{s=1}^\infty z^{-s} \frac{(1-t^{-s})}{s q^{-s}}
 \sum_{i=1}^N\lambda_{i}^{-s} \right)
 \right\rangle_{\tau}^{N_f}
\end{aligned}
\end{equation}
where we recall the definition \eqref{eq:mass_powersum} of the power-sum variable $p_{s}(u)$, while
\begin{equation}
\label{eq:LHS_new}
\begin{aligned}
 (\mathrm{LHS}) & = \lbracket\sum_{n\in\mathbb{Z}}z^n\sum_{i=1}^N
  \lambda_{i}^n G_{i}(\lambda ;t) \rbracket_\tau^{N_f}\\ 
 & =
 \frac{1}{1-t}\left\langle\exp\left(\sum_{s=1}^\infty z^{-s} \frac{(1-t^s)}{s}
 \sum_{i=1}^N\lambda_{i}^{-s} \right)\right\rangle_{\tau}^{N_f}
 -\frac{t^N}{1-t}\left\langle\exp\left(\sum_{s=1}^\infty z^{s} \frac{(1-t^{-s})}{s}
 \sum_{i=1}^N\lambda_{i}^{s} \right)
 \right\rangle_{\tau}^{N_f} ~.
\end{aligned}
\end{equation}
For order $n\geq 0$ in the variable $z$ all the insertions contain only positive powers of $\lambda_{i}$, therefore the corresponding constraints can be expressed as PDEs in the time variables.
At order $n=-1$ in $z$, there are negative power contributions in $\lambda_{i}$ which must vanish for the constraint to be meaningful. The offending term is
\begin{equation}
 \frac{1-q^{1-\kappa_1} t^{N-1}}{1-t}
 z^{-1}
 \lbracket \sum_{i=1}^N \lambda_{i}^{-1}\rbracket_\tau^{N_f}~,
\end{equation}
which vanishes precisely if we let $t=q^{\beta}$ with $\beta \in \mathbb{C}$ and impose the {\it balancing condition}
\begin{equation}
\label{eq:balancing_1}
 \kappa_1 = \beta(N-1)+1~,
\end{equation}
which is consistent with the choice of FI parameter in \cite{Nedelin:2016gwu}.
More generally we define 
\begin{equation}
\label{eq:balancingD2S1}
 \dvar := \kappa_1 -\beta(N-1)-1~, \qquad \qquad r := q^{\dvar}~
\end{equation}
and we refer to $\dvar$ as the \textit{balancing parameter}, which vanishes exactly when the balancing condition is satisfied.
The $q$-Virasoro constraints can then be written as
\begin{equation}
\label{eq:qVirasoroD2S1}
\begin{aligned}
 & \frac{t^{N}}{1-t}
 \exp\left(\sum_{s=1}^\infty z^{s}\frac{(1-t^{-s})}{s}\partial_{s}\right)
 \Z_{\halfindex}^{N_f}(\tau) + \\
& \qquad + \frac{r^{-1} q^{-1} t^{1-N}}{1-t}
 \exp\left( \sum_{s=1}^\infty z^{-s}(1-q^s)\left(\tau_{s}
 +\frac{p_{s}(u)}{s(1-q^{-s})}\right) \right)
 \exp\left(\sum_{s=1}^\infty z^{s} \frac{(1-t^s)}{s q^s}
 \partial_{s} \right)\Z_{\halfindex}^{N_f}(\tau) = \\
 & =\frac{1}{1-t}
 \left\langle
 \exp\left(\sum_{s=1}^\infty z^{-s} \frac{(1-t^s)}{s}
 \sum_{i=1}^N\lambda_{i}^{-s} \right)\right\rangle_{\tau}^{N_f}+ \\
& \qquad  + \frac{r^{-1} q^{-1} t}{1-t}
 \exp\left( \sum_{s=1}^\infty z^{-s}(1-q^s)\left(\tau_{s}
 +\frac{p_{s}(u)}{s(1-q^{-s})}\right) \right)
 \left\langle\exp\left(\sum_{s=1}^\infty z^{-s} \frac{(1-t^{-s})}{s q^{-s}}
 \sum_{i=1}^N\lambda_{i}^{-s} \right)
 \right\rangle_{\tau}^{N_f}~.
\end{aligned}
\end{equation}
The precise relation between these constraints and the generators of the $q$-Virasoro algebra are delucidated in Appendix~\ref{sec:generators_relation}.
\\
\\
For the purpose of actually writing these constraints as recursion relations for the correlation functions $c_\rho$, it is more convenient to rewrite the shift of times as multiplication by a polynomial in $z^{-1}$ and the masses $u_k$, using the identity
\begin{equation}
\label{eq:coeffAk}
 \exp\left(-\sum_{s=1}^\infty {\left(\frac{q}{z}\right)^s} \frac{p_{s}(u)}{s} \right) =
 \prod_{k=1}^{N_f} \left(1-\frac{q}{z} u_k \right) =
 \sum_{k=0}^{N_f}A_k\left(\frac{q}{z}\right)^k~,
\end{equation}
where the coefficients $A_k=A_k(u)$ are defined as antisymmetric Schur polynomials in the masses $u_k$,
\begin{equation}
 A_k = (-1)^k\schur_{\{\underbrace{1,\dots,1}_{k}\}}(p_s=p_s(u))~.
\end{equation}
Notice that antisymmetric Schur polynomials of degree higher than the number of variables $\{u_k\}$ are identically zero, therefore we have only a finite number of coefficients $A_k$ and the series expansion in \eqref{eq:coeffAk} is truncated to degree $N_f$.

\subsubsection{$S^3_b$}

Let us now derive the constraints for the $S^3_b$ generating function \eqref{eq:generatingfunction}.
As already mentioned, there are two sets of independent $q$-Virasoro constraints that we can write.
They correspond to the Ward identities obtained by acting with $q$-difference operators that shift each set of variables $\{\lambda_{i,\alpha}\}$ separately. Namely, we can define the operators $\shift$ which act as:
\begin{equation}
\label{eq:shift_operator}
\begin{aligned}
  \hat{\mathsf{M}}_{i,1} f(X) = & f (\ldots, X_i - \omega_{2}, \ldots) \\
  \hat{\mathsf{M}}_{i,2} f(X) = & f (\ldots, X_i - \omega_{1}, \ldots)~.
\end{aligned}
\end{equation}
Because of the periodicity of the exponential function, it follows that on the exponentiated variables $\{\lambda_{i,\alpha}\}$ the shift acts multiplicatively as
\begin{equation}
  \hat{\mathsf{M}}_{i,\alpha} \lambda_{j,\alpha'} = \left\{
\begin{array}{ll}
 q_\alpha^{-1}\lambda_{j,\alpha'} & \text{if $(i,\alpha)=(j,\alpha')$} \\
 \lambda_{j,\alpha'} & \text{otherwise}
\end{array}
\right.
\end{equation}
which explains why we use the same symbol as in \eqref{eq:shiftoperator}.
Proceeding as above, we define functions $G_{i, \alpha}(\lambda ;t_\alpha)$ as
\begin{equation}
 G_{i,\alpha}(\lambda ;t_\alpha) = \prod_{j\neq i}\frac{1-t_{\alpha}\lambda_{i,\alpha}/\lambda_{j,
  \alpha}}{1-\lambda_{i,\alpha}/\lambda_{j,\alpha}}~,
\end{equation}
and compute the Ward identities as the integral equations
\begin{equation}
\label{eq:qDifferenceOperator}
 \int_{(\mathi\mathbb{R})^N} \prod_{i=1}^N \mathd X_{i}\,
 \sum_{i=1}^N(\hat{\mathsf{M}}_{i,\alpha}-1)\left[\lambda_{i,\alpha}^{n}
 G_{i,\alpha}(\lambda ;t_\alpha)\dots \right]
 = 0~.
\end{equation}
Keeping $\alpha$ fixed, we sum over all $n\in\mathbb{Z}$ with weight $z^n$ and obtain the $q$-Virasoro constraint
\begin{equation}
\label{eq:qVirasoroS3}
\begin{aligned}
 & t^{N}_\alpha
 \exp\left(\sum_{s=1}^\infty z^{s}\frac{(1-t^{-s}_\alpha)}{s}\partial_{s,\alpha}\right)
 \Z_{S^3_b}^{N_f}(\tau) + \\
 & \qquad + r_\alpha^{-1} q^{-1}_\alpha t^{1-N}_\alpha
 \exp\left( \sum_{s=1}^\infty z^{-s}(1-q_\alpha^s)\left(\tau_{s,\alpha}
 +\frac{p_{s}(u)}{s(1-q^{-s}_\alpha)}\right) \right)
 \exp\left(\sum_{s=1}^\infty z^{s} \frac{(1-t^s_\alpha)}{s q^s_\alpha}
 \partial_{s,\alpha} \right)\Z_{S^3_b}^{N_f}(\tau) = \\
 & = 
 \left\langle
 \exp\left(\sum_{s=1}^\infty z^{-s} \frac{(1-t^s_\alpha)}{s}
 \sum_{i=1}^N\lambda_{i,\alpha}^{-s} \right)\right\rangle_{\tau}^{N_f} + \\
 & \qquad + r_\alpha^{-1} q^{-1}_\alpha t_\alpha
 \exp\left( \sum_{s=1}^\infty z^{-s}(1-q_\alpha^s)\left(\tau_{s,\alpha}
 +\frac{p_{s}(u)}{s(1-q^{-s}_\alpha)}\right) \right)
 \left\langle\exp\left(\sum_{s=1}^\infty z^{-s} \frac{(1-t^{-s}_\alpha)}{s q_\alpha^{-s}}
 \sum_{i=1}^N\lambda_{i,\alpha}^{-s} \right)
 \right\rangle_{\tau}^{N_f}~,
\end{aligned}
\end{equation}
where we defined the balancing parameters $\dvar$ and $r_\alpha$ as
\begin{equation}
 \label{eq:balancingparameterS3}
 \omega\dvar := \kappa_1 - \omega - M_{\mathrm{a}}(N-1) + \frac{\omega}{2}\frac{N_f}{2}
 + \sum_{k=1}^{N_f}\frac{m_k}{2}~,
 \quad\quad\quad
 r_\alpha := \mathe^{\frac{2\pi\mathi\omega}{\omega_\alpha}\dvar} = q_\alpha^\dvar~.
\end{equation}
If we explicitly expand \eqref{eq:qVirasoroS3} in powers of $z$, we have that for $n\geq0$ all equations are free of expectation values of negative powers of $\{\lambda_{i,\alpha}\}$ and therefore can be rewritten as PDEs for the generating function. At order $n=-1$, there are negative power contributions which must be cancelled. An explicit computation shows that these offending terms vanish exactly when $\dvar=0$ (so that $r_\alpha=1$). We call this the \textit{balancing condition}. Observe that $\dvar$ corresponds to the combination $\sqrt{\beta}\alpha$ of \cite{Nedelin:2016gwu}.
\\
\\
Finally, by applying the reasoning of \cite{Cassia:2019sjk}, we find that the correlators $c_{\rho;\rho'}$ in \eqref{eq:s3_correlators} obtained as solutions to the above constraints, do factorize according to
\begin{equation}
\label{eq:factorizationcorr}
 c_{\rho;\rho'} = \frac{c_{\rho;\emptyset}\cdot c_{\emptyset;\rho'}}{c_{\emptyset;\emptyset}}~,
\end{equation}
where $c_{\rho;\emptyset}$ is a correlation function containing only the $\{\lambda_{i,1}\}$, and $c_{\emptyset;\rho'}$ is a correlation function only containing the $\{\lambda_{i,2}\}$.
The factorization of the correlators then implies that \eqref{eq:formalseriesS3} can be written as a product
\begin{equation}
\label{eq:seriesfactorization}
 \Z_{S_b^3}^{N_f}(\tau) = \Z_{S_b^3}^{N_f}(0)
 \left[\sum_{\rho}\frac{1}{|\Aut(\rho)|}\frac{c_{\rho;\emptyset}}{c_{\emptyset;\emptyset}}
 \prod_{a\in\rho}\tau_{a,1}\right]
 \cdot
 \left[\sum_{\rho}\frac{1}{|\Aut(\rho)|}\frac{c_{\emptyset;\rho}}{c_{\emptyset;\emptyset}}
 \prod_{a\in\rho}\tau_{a,2}\right]
\end{equation}
where each half is a formal power series in one set of times only, and it satisfies its own copy of the $q$-Virasoro constraints. Here we also used that the empty correlator is the partition function, i.e. the generating function evaluated on all the times equal to zero, $c_{\emptyset;\emptyset} \equiv \Z_{S_b^3}^{N_f}(0)$. Using this fact, we deduce that up to normalization, the correlators of the $S^3_b$ theory are products of correlators of $\halfindex$ theories. The formula \eqref{eq:seriesfactorization} gives a precise meaning to the equivalence proposed in \eqref{eq:S3andD2S1}.
\\
\\
\textbf{Remark}. 
As observed in the pole analysis of the partition function in \cite[Appendix~C]{Cassia:2019sjk}, the choice of shift operator $\shift$ is motivated by the requirement that when we shift the contour of integration in the LHS of the constraint equation in order to reabsorb the action of the shift in \eqref{eq:shift_operator}, we should not cross any poles of the integrand.
Since the only poles that can come in between the two contours are the ones due to the fundamental matter fields, the choice of the chirality of these fields is restricted by the sign of the shift in the variables. Namely, for negative shift of the $X_i$ variables, we need to shift the contour to the left and therefore we cannot have any poles to the left of the imaginary axis. This implies that all fundamental matter fields should be anti-chiral rather than chiral. Their precise number is then fixed by the value of the bare CS level as in \eqref{eq:effectiveCS}.
However, if we were to use the opposite shift operator $\shift^{-1}$, we would need to shift the contour in the opposite direction, and by doing so we might cross some poles along the way. To avoid this problem we can substitute all anti-chiral fundamental multiplets with chiral fundamentals. The change of chirality inverts the position of the poles of the corresponding double sine functions thus allowing us to shift the contour.
By reproducing the analysis of the $q$-Virasoro constraints we find that the corresponding condition on the number of fundamental chiral fields is $N_f=-2\kappa_2$, which can only be satisfied for negative bare CS level. The rest of the derivation is completely equivalent provided we adopt the substitution of parameters
\begin{equation}
\label{eq:change_of_parameters}
 q_\alpha \mapsto q_\alpha^{-1}, \quad
 t_\alpha \mapsto t_\alpha^{-1}, \quad
 r_\alpha \mapsto r_\alpha^{-1}, \quad
 u_{k,\alpha} \mapsto u_{k,\alpha}^{-1}.
\end{equation}
This is indeed compatible with the symmetry properties of the partition function
proved in \cite[Proposition 5.3.16]{vdBult:2007}.

\subsection{Solution of the constraints}
\label{sec:quantum_recursive_solution}

We now study the conditions under which the $q$-Virasoro constraint equations admit solutions and whether these solutions are unique or not. As we shall show, the only cases in which the solution is uniquely defined are those of $N_f=1,2$ and in such cases we use the solution (computed algorithmically up to a certain finite order) to explicitly verify the property of averages of Macdonald polynomials conjectured in \cite{Morozov:2018eiq}. Having found an explicit solution, we study the semi-classical limit and propose an identification with the models of Section~\ref{sec:classical_models}.
\\
\\
Before actually solving the constraints \eqref{eq:qVirasoroD2S1} and \eqref{eq:qVirasoroS3}, we observe that because of the factorization property \eqref{eq:seriesfactorization}, we just need to find a solution to the $q$-Virasoro constraints on $\halfindex$. Up to normalization then, a solution for the generating function on $S^3_b$ can be obtained by taking the product of correlation functions as in \eqref{eq:factorizationcorr}. For this reason in this section
we consider an \textit{abstract generating function} $\Z^{N_f}(\tau)$ which depends on a single set of times and satisfies one set of $q$-Virasoro constraints with respect to those times. 
\\
\\
The first step in finding a solution to \eqref{eq:qVirasoroD2S1} is to rewrite the whole set of constraints as an equation
\begin{equation}
\label{eq:qtConstr}
 \constr_{q,t}\, \Z^{N_f}(\tau) = 0~, 
\end{equation}	
where $\constr_{q,t}$ is the operator obtained by appropriately re-summing the individual constraints, similarly to how we did in the classical case. What we then find is that $\constr_{q,t}$ is a quantum deformation of the operators in \eqref{eq:p1classicalconstr} and \eqref{eq:p2classicalconstr}. Moreover, we observe that the $q$-deformation preserves the triangular form\footnote{The main algebraic difference between $\constr_{q,t}$ and $\constr_{\mathrm{classical}}$ is that the former is only triangular with respect to the full basis of monomials with the total ordering induced by integer partitions, while the latter is also triangular with respect to the partial order given by the monomial degree (size of the partition). } of the operator.
In order to see this, we manipulate \eqref{eq:qVirasoroD2S1} by separating the exponential of the times from the exponential of the $p_s(u)$ and then applying \eqref{eq:coeffAk} to obtain
\begin{equation}
\label{eq:qVirasoroD2S1_rearranged}
\begin{aligned}
 & \frac{t^{N}}{1-t}
 \exp\left(-\sum_{s=1}^{{\infty}}z^{-s}\left(1-q^{s}\right){\tau}_{s}\right)
 \exp\left(\sum_{s=1}^\infty z^{s}\frac{(1-t^{-s})}{s}\partial_{s}\right)
 \Z^{N_f}(\tau) \\
 & + \frac{r^{-1} q^{-1} t^{1-N}}{1-t}
 \prod_{k=1}^{N_f} \left(1-q u_{k}z^{-1} \right)
 \exp\left(\sum_{s=1}^\infty z^{s} \frac{(1-t^s)}{s q^s}
 \partial_{s} \right)\Z^{N_f}(\tau) = \\
 & = \frac{1}{1-t}
 \exp\left(-\sum_{s=1}^{{\infty}}z^{-s}\left(1-q^{s}\right){\tau}_{s}\right)
 \left\langle
 \exp\left(\sum_{s=1}^\infty z^{-s} \frac{(1-t^s)}{s}
 \sum_{i=1}^N\lambda_{i}^{-s} \right)\right\rangle_{\tau}^{N_f} \\
 & + \frac{r^{-1} q^{-1} t}{1-t}
 \prod_{k=1}^{N_f} \left(1-q u_{k}z^{-1} \right)
 \left\langle\exp\left(\sum_{s=1}^\infty z^{-s} \frac{(1-t^{-s})}{s q^{-s}}
 \sum_{i=1}^N\lambda_{i}^{-s} \right)
 \right\rangle_{\tau}^{N_f} ~.
\end{aligned}
\end{equation}
The operator $\constr_{q,t}$ is defined by expanding in powers of $z$ and re-summing over all $n\geq-1$ with weight $(n+N_f)\tau_{n+N_f}$. Finally, one can check by explicitly writing $\constr_{q,t}$ as a linear operator on the formal ring $\mathbb{C}[[\tau_1,\tau_2,\dots]]$, that it assumes the form of a semi-infinite triangular matrix\footnote{Observe that every triangular system of linear equations is equivalent to a set of recursion relations between the components of the solution. In our case it corresponds to an infinite set of finite-step recursion relations between the correlators $c_\rho$.}.
\\
\\
If we split the constraint operator as
\begin{equation}
\label{eq:D-W}
 \constr_{q,t} = D_{q,t} - W_{q,t}
\end{equation}
where $D_{q,t}$ is the diagonal part and $W_{q,t}$ is the strictly upper triangular, then as we will show in the following sections, we can write
\begin{equation}
 D_{q,t} \approx
 \sum_{s=N_f-1}^\infty \left(\frac{1-t^{s}}{1-t}q^{-s}\right)
 \tau_{s}\partial_{s} ~,
\end{equation}
up to some overall coefficient which does not depend on the times.
For $N_f=1,2$ the operator is diagonal with all eigenvalues different from zero except for the one associated to the constant monomial (labeled by the empty partition), hence we deduce that both $D_{q,t}$ and $\constr_{q,t}$ have a 1-dimensional kernel just like in the classical case of $\p=1,2$. This means that the solution to the equation \eqref{eq:qtConstr} is unique up to normalization of the empty correlator $c_\emptyset$. For higher $N_f$ the kernel becomes infinite dimensional and the solution is no longer uniquely defined.
This suggest a strong parallel with the classical models upon identification of the parameters $N_f$ and $\p$. In the following sections we analyze this question in detail and provide a concrete limiting procedure which explicitly shows the correspondence.

\subsubsection{$N_f=1$}

We first consider the case of $N_f=1$. Using Cauchy's formula \eqref{eq:symm-schur-poly} we can expand \eqref{eq:qVirasoroD2S1_rearranged} in powers $z^n$ for $n\geq0$ to get
\begin{equation}
\label{eq:modified_combined_constraint}
\begin{split}
 & \frac{t^{N}}{1-t} \sum_{\ell=0}^\infty
 \schur_{\{\ell\}}(p_s = - s\left(1-q^{s}\right){\tau}_{s})
 \schur_{\{\ell+n\}}\left(p_s = \left(1-t^{-s}\right)
 \partial_{s}\right)
 \Z^{N_f = 1}(\tau)+\\
 & + \frac{r^{-1} q^{-1} t^{1-N}}{1-t}
 \sum_{k=0,1} q^k A_{k}
 \schur_{\{n+k\}}
 \left(p_s=\frac{1-t^{s}}{q^s}\partial_{s}\right)
 \Z^{N_f=1}(\tau)
 - \delta_{n,0} \frac{(1+r^{-1}q^{-1}t)}{1-t} \Z^{N_f = 1}(\tau)
 = 0~.
\end{split}
\end{equation}
If we define the differential operator $\constr_n$ as the operator such that
$\constr_n\Z^{N_f = 1}(\tau)=0$ is the $n$-th $q$-Virasoro constraint \eqref{eq:modified_combined_constraint}, then we re-sum all such operators over $n\geq0$ to define
\begin{equation}
 \constr_{q,t} := \sum_{n=0}^\infty (n+1) \tau_{n+1} \constr_n~,
\end{equation}
which we can write explicitly as
\begin{equation}
\label{eq:operatorNf1}
\begin{aligned}
 \constr_{q,t} = & \frac{r^{-1} t^{1-N}}{1-t} A_{1}
 \sum_{n=0}^\infty n \tau_{n} \schur_{\{n\}}
 \left(p_s=\frac{1-t^{s}}{q^s}\partial_{s}\right) + \\
 & + \frac{r^{-1} q^{-1} t}{1-t}
 \left[ t^{-N} \sum_{n=0}^\infty (n+1) \tau_{n+1} \schur_{\{n\}}
 \left(p_s=\frac{1-t^{s}}{q^s}\partial_{s}\right)
 -\tau_{1}\right] + \\
 & + \frac{1}{1-t}\left[t^N \sum_{\ell,n=0}^\infty (n+1)\tau_{n+1}
 \schur_{\{\ell\}}(p_s = - s\left(1-q^{s}\right){\tau}_{s})
 \schur_{\{\ell+n\}}\left(p_s = \left(1-t^{-s}\right)
 \partial_{s} \right) -\tau_1\right]
\end{aligned}
\end{equation}
where the first term in the right hand side is an operator of degree 0 while the other two are of degree 1. The degree zero part is the one that contains the diagonal operator $D_{q,t}$ which is defined by expanding the symmetric Schur as in \eqref{eq:symmschur} and picking only the term $p_n/n$ (all other terms are higher derivatives in the times and therefore cannot be diagonal),
\begin{equation}
 D_{q,t} = {r^{-1} t^{1-N}} A_{1}
 \sum_{n=0}^\infty \tau_{n}
 \frac{1-t^{n}}{1-t}q^{-n}\partial_{n}~.
\end{equation}
The remaining terms in the r.h.s. of \eqref{eq:operatorNf1} give the definition of $-W_{q,t}$.
\\
\\
Now we have that the full set of $q$-Virasoro constraints is equivalent to the statement that $\Z^{N_f = 1}(\tau)$ is in the kernel of the operator $\constr_{q,t}$. The fact that this operator is not homogeneous in degree then corresponds to a set of recursion relations between correlators in degree $d$ and those in degree $d-1$. In particular, the recursion will allow us to write $c_\rho$ as a linear combination of all $c_{\rho'}$ such that $\deg(\rho')\geq\deg(\rho)-1$ and $\rho'<\rho$. Because the linear operator in \eqref{eq:operatorNf1} has a 1-dimensional kernel, the generating function $\Z^{N_f = 1}(\tau)$ is unique up to a choice of a normalization constant $c_{{\emptyset}}\equiv\Z^{N_f = 1}(0)$.
An exact solution up to degree 3 is then given by the correlators
\begin{equation}
\label{eq:correlators_nf1}
\begin{aligned}
 c_{\{3\}} & = \frac{\left(t^{{N}}-1\right) \left(q r t^{{N}}-t\right) }{A_1^3 t^6 \left(t^3-1\right)} \Bigg(q^2 r^2 t^{4 {N}} \left(q \left(t^2+t+1\right) \left(q^2 t+q+1\right)+1\right)+t^{2 {N}+3} \left(q^2 r^2+q r t+t^2\right)+\\
 & + t^{{N}+4} (q r+t)-q r t^{3 {N}+1} \left(q (q+1) \left(t^2+t+1\right)+t+1\right) (q r+t)+t^5\Bigg) c_{{\emptyset}},\\
 c_{\{2,1\}} & = \frac{\left(t^{{N}}-1\right) \left(q r t^{{N}}-t\right) }{A_1^3 (t-1)^2 t^6 (t+1)}\Bigg(q^2 r^2 t^{4 {N}} (q (t+1) (q (t-1) (q t+1)+t)+1)+\\
 & + t^{2 {N}+3} \left(q^2 r (r+t+1)+q r (t+1)+t^2\right)-q (q+1) r (t+1) t^{3 {N}+1} (q (t-1)+1) (q r+t)-t^5\Bigg)c_{{\emptyset}},\\
 c_{\{1,1,1\}} & = \frac{\left(t^{{N}}-1\right) \left(q r t^{{N}}-t\right) }{A_1^3 (t-1)^3 t^6}\Bigg(q^2 r^2 t^{4 {N}} (q (t-1) (q (t-1) (q t+1)+t+2)+1)+\\
 & + t^{2 {N}+3} \left(q^2 r (r+3 t-3)+q r (t+3)+t^2\right)-2 t^{{N}+4} (q r+t)+\\
 & -q r t^{3 {N}+1} (q (t-1)+1) (q (t-1)+t+1) (q r+t)+t^5\Bigg)c_{{\emptyset}},\\
 c_{\{2\}} & = \frac{\left(t^{{N}}-1\right) \left(q r t^{{N}}-t\right) \left(q r t^{2 {N}} (q t+q+1)-t^{{N}+1} (q r+t)-t^2\right)}{A_1^2 t^3 \left(t^2-1\right)} c_{{\emptyset}},\\
 c_{\{1,1\}} & = \frac{\left(t^{{N}}-1\right) \left(q r t^{{N}}-t\right) \left(q r t^{2 {N}} (q (t-1)+1)-t^{{N}+1} (q r+t)+t^2\right)}{A_1^2 (t-1)^2 t^3}c_{{\emptyset}},\\
 c_{\{1\}} & = \frac{\left(t^{{N}}-1\right) \left(q r t^{{N}}-t\right)}{A_1 (t-1) t} c_{{\emptyset}},
\end{aligned}
\end{equation}
which are rational functions in the parameters $q$, $t$, $r$ and $A_1$. Moreover, we have that all correlators of degree higher than 1 are proportional to $c_{\{1\}}$ precisely as in \eqref{eq:classical_correlators_p1}.
\\
\\
Observe that the dependence on the rank of the gauge group $N$ comes only through the powers of $t$, and in the large $N$ limit the correlators behave as
\begin{equation}
 c_{\rho}/c_{\emptyset} \sim t^{2N|\rho|}~,
\end{equation}
where $|\rho|$ is the size of the corresponding partition.

\subsubsection*{Semi-classical limit}

As we already observed, the case of $N_f=1$ bears many similarities with the classical model with $\p=1$. The precise relation can be established by a semi-classical limit procedure in which we introduce a perturbation parameter $\hbar$ such that $q=\mathe^{\hbar}$. In the limit $\hbar\to0$ both the constraint equations and their solution match exactly with the formulas for the classical case $\p=1$. The matching goes as follows.
\\
\\
First we consider the measure $\Delta_{q,t}$ in the generating function. Assuming $t=q^\beta$ with $\beta\in\mathbb{Z}$, we can write
\begin{equation}
\label{eq:qtmeasurerewrite}
 \Delta_{q,t}(\lambda) = \prod_{1\leq i\neq j\leq N} \prod_{n=0}^{\beta-1}(1-q^n\lambda_i/\lambda_j)
 = \prod_{1\leq i\neq j\leq N} \prod_{n=0}^{\beta-1}(\lambda_i-q^n\lambda_i)
 \prod_{i=1}^N \lambda_i^{-\beta(N-1)}
\end{equation}
which in the limit $\hbar\to0$ (or $q\to1$) clearly gives a $\beta$-deformed Vandermonde determinant $\Delta^{2\beta}(\lambda)$ multiplied by a determinant insertion of power $-\beta(N-1)$. Combining this determinant with the one coming from the FI term and the one in the measure $\mathd\lambda_i/\lambda_i$, we have
\begin{equation}
 \prod_{i=1}^N \lambda_{i}^{\dvar}~,
\end{equation}
where we used \eqref{eq:balancingD2S1} to write the exponent as the balancing parameter $\dvar$.
\\
\\
The last terms to match are the 1-loop determinant of the fundamental anti-chiral and the potential $V(\lambda_i)$. From \eqref{eq:shiftoftimesD2S1other} we see that the shift of times is singular in the limit $q\to1$. In order to make this limit well defined, we also scale the mass $u_{1}$ as
\begin{equation}
\label{eq:massesNf1}
 u_{1} = q^{-1}(1-q) a_1
 \quad\Rightarrow\quad
 A_1 = a_1\hbar + O(\hbar^2),
\end{equation}
with $a_1$ a positive constant independent of $\hbar$. This way we can re-write the 1-loop determinant as
\begin{equation}
 \prod_{i=1}^N (q\lambda_i u_1;q)_\infty
 = \exp\left(-\sum_{s=1}^\infty\frac{(1-q)^s}{1-q^s}\frac{a_1^s}{s}\sum_{i=1}^N\lambda_i^s\right)
\end{equation}
which corresponds to the shift of times
\begin{equation}
 \tau_s\mapsto\tau_s-\frac{(1-q)^s}{1-q^s}\frac{a_1^s}{s}
\end{equation}
and in the limit $q\to1$ the shift becomes
\begin{equation}
 \lim_{q\to1}\frac{(1-q)^s}{1-q^s}\frac{a_1^s}{s} = a_1\delta_{s,1}~.
\end{equation}
Hence we obtain the same integrand as in \eqref{eq:WishartLaguerre}. Notice that in order to get the correct Vandermonde term in the semi-classical limit, we had to assume that $\beta$ is an integer number. For arbitrary complex values however \eqref{eq:qtmeasurerewrite} does not make sense, nevertheless we know that the generating function admits an analytic continuation in $\beta$, then we can assume that our limit still makes sense even for non-integer complex numbers.
\\
\\
With the choice of mass as in \eqref{eq:massesNf1} the semi-classical limit $q\to1$ is also well-defined at the level of the $q$-Virasoro constraints. To this end we assume the following power series expansion in the parameter $\hbar$ around 0 both for the generating function and for the constraint operators themselves,
\begin{equation}
\label{eq:hbarexpansionZ}
\begin{aligned}
 \Z^{N_f} & = \Z^{N_f}_{(0)} + \hbar \Z^{N_f}_{(1)} + \hbar^2 \Z^{N_f}_{(2)} + \dots \\
 \constr_n &= \constr^{(0)}_n + \hbar \constr^{(1)}_n + \hbar^2 \constr^{(2)}_n + \dots \\
\end{aligned}
\end{equation}
where one can check that $\constr_n^{(0)}=0$. The constraint equation \eqref{eq:modified_combined_constraint} can be expanded as
\begin{equation}
 \label{eq:hbarexpansionconstr}
 \constr_n \Z^{N_f} =
 \hbar \left( \constr^{(1)}_n\Z^{N_f}_{(0)} \right) +
 \hbar^2 \left( \constr^{(2)}_n\Z^{N_f}_{(0)} + \constr^{(1)}_n\Z^{N_f}_{(1)} \right)
 + O(\hbar^3) = 0
\end{equation}
where the first non-trivial equation comes by setting the coefficient of $\hbar$ to zero.
We thus have that $\Z_{(0)}^{N_f=1}$ is in the kernel of $\constr^{(1)}_n$, with
\begin{equation}
\begin{aligned}
 -\constr^{(1)}_n =
 \bigg[ \beta \sum_{a+b=n} \partial_a\partial_b 
 & + \left((1-\beta)(n+1)+\dvar + 2\beta N \right) \partial_n + \\
 & + \sum_{s=1}^\infty s \tau_s \partial_{s+n}
 + \delta_{n,0}N\left(\dvar + \beta(N-1)+1\right) -a_1\partial_{n+1}
 \bigg]~.
\end{aligned}
\end{equation}
This is precisely the operator in \eqref{eq:semiclassicalvirasoroNf1}. Hence we have the identification of generating functions
\begin{equation}
 \Z_{(0)}^{N_f=1}(\tau) \equiv \Z^{\p=1}(\tau)~,
\end{equation}
and we can check explicitly that the result \eqref{eq:correlators_nf1} is consistent with \eqref{eq:classical_correlators_p1} in the limit $\hbar\to0$. Finally, we notice that the quantum operator $D_{q,t}$ satisfies
\begin{equation}
 D_{q,t} = \hbar \, a_1 D + O(\hbar^2)~.
\end{equation}
We remark that in the case of the generating function on $S^3_b$ there are two sets of exponentiated masses $u_{1,\alpha}$ but only one actual fundamental mass $m_1$. Therefore our parametrization in \eqref{eq:massesNf1} leads to a well-defined limit only for one copy of $q$-Virasoro at a given time. Namely, if the mass has been chosen so that one copy of the constraints becomes usual Virasoro in the semi-classical limit, then the other copy is not well-behaved in that limit.

\subsubsection*{Averages of Macdonald Polynomials}

From the point of view of the $q$-deformed matrix models it is natural to consider expectation values of Macdonald polynomials. These polynomials are the natural quantum generalization of the Schur polynomials and as such they can be interpreted as the correct quantum characters of the model. We now show that their averages do indeed satisfy a special property of the form $\langle\mathsf{character}\rangle=\mathsf{character}$.
\\
\\
Introducing the functions
\begin{equation}
\label{eq:pikNdef}
 \pi^{(N)}_k := \frac{t^{\frac{k}{2}N}-t^{-\frac{k}{2}N}}{t^{\frac{k}{2}}-t^{-\frac{k}{2}}},
 \quad\quad\quad
 \hat{\pi}^{(N)}_k := r^{\frac{k}{2}} q^{\frac{k}{2}} t^{k(N-1)}
 \frac{r^{\frac{k}{2}} q^{\frac{k}{2}} t^{\frac{k}{2}(N-1)}
  - r^{-\frac{k}{2}} q^{-\frac{k}{2}} t^{-\frac{k}{2}(N-1)}}
 {t^{\frac{k}{2}}-t^{-\frac{k}{2}}}
\end{equation}
and
\begin{equation}
 \delta^\ast_{k,1} := \frac{u_1^{k}} {t^{-\frac{k}{2}}-t^{\frac{k}{2}}}
 =(-1)^{k+1}q^{-\frac{k}{2}}\frac{(q^{\frac12}-q^{-\frac12})^k}{t^{\frac{k}{2}}-t^{-\frac{k}{2}}}a_{1}^k~,
\end{equation}
we have the identity
\begin{equation}
\label{eq:macdonaldNf1}
 \lbracket\macdonald_\rho(p_k)\rbracket^{N_f=1} =
 \frac{\macdonald_\rho(p_k=\hat{\pi}^{(N)}_{k})}
 {\macdonald_\rho(p_k=\delta^\ast_{k,1})}\,
 \macdonald_\rho(p_k=\pi^{(N)}_{k}) \, c_{\emptyset}~.
\end{equation}
We have checked that this formula holds for all partitions up to degree 6. In principle our solution should allow to check up to arbitrary finite order, however for higher degrees the computation becomes quickly too impractical even for computer calculations.
\\
\\
In the semi-classical limit $t=q^\beta$, $r=q^\dvar$,
$u_1=-(1-q^{-1})a_1$ and $q\to1$, Macdonald polynomials degenerate to Jack
polynomials $\jack_\rho(p_k)$ and the formula \eqref{eq:macdonaldNf1} coincides with \eqref{eq:jackNf1}. Furthermore, we can set $\beta = 1$ to recover the average of Schur polynomials in \eqref{eq:schurNf1}.

\subsubsection{$N_f=2$}

In order to show that a solution in this case exists and is unique we need to use all constraints for $n\geq -1$ and in particular we need the $n=-1$ constraint to be well-defined. This implies the balancing condition $r=1$, (i.e. $\dvar=0$). The $n$-th $q$-Virasoro constraint operator $\constr_n$ can be written as
\begin{equation}
\begin{split}
 \constr_n = & \frac{t^{N}}{1-t} \sum_{\ell=0}^\infty
 \schur_{\{\ell\}}(p_s = - s\left(1-q^{s}\right){\tau}_{s})\,
 \schur_{\{\ell+n\}}\left(p_s = \left(1-t^{-s}\right)
 \partial_{s} \right) + \\
 & + \frac{q^{-1} t^{1-N}}{1-t}
 \sum_{k=0}^{N_f} q^k A_{k} \schur_{\{n+k\}}
 \left(p_s=\frac{1-t^{s}}{q^s}\partial_{s}\right)
 - \delta_{n,0} \frac{1+q^{-1}t}{1-t} \\
 & + \delta_{n,-1} \left[\frac{1-q}{1-t}\tau_{1} + \frac{t}{1-t} A_{1} \right]
\end{split}
\end{equation}
so that the quantum operator $\constr_{q,t}=\sum_{n=-1}^\infty (n+2)\tau_{n+2}\constr_n$ takes the form
\begin{equation}
\label{eq:qConstrOperatorNf2}
\begin{aligned}
 \constr_{q,t} =
 & \frac{t^{1-N}}{1-t}A_2 q \sum_{s=1}^\infty s \tau_s \schur_{\{s\}}\left(p_k=\frac{1-t^k}{ q^k}\partial_k \right) + \\
 & + \frac{t^{1-N}}{1-t}A_1 \sum_{s=1}^\infty s \tau_s \schur_{\{s-1\}}\left(p_k=\frac{1-t^k}{q^k}\partial_k \right) - \frac{t}{1-t} A_1\tau_1 + \\
 & + \frac{t^{1-N}}{1-t}q^{-1} \sum_{s=1}^\infty s \tau_s \schur_{\{s-2\}}\left(p_k=\frac{1-t^k}{q^k}\partial_k \right) - \frac{q^{-1}t}{1-t}2\tau_2 + \\
 & + \frac{t^{N}}{1-t}\sum_{\ell=0}^\infty\sum_{s=1}^\infty s \tau_s \schur_{\{\ell\}}\left(p_k= -k(1-q^k)\tau_k \right)\schur_{\{\ell+s-2\}}\left(p_k= {(1-t^{-k})}\partial_k \right) +\\
 & - \frac{1}{1-t}2\tau_2 + \frac{1-q}{1-t}\tau_1^2
\end{aligned}
\end{equation}
where the operator in the first line is of degree 0, the one in the second line is of degree 1 and those in the other three lines are of degree 2.
\\
\\
Similarly to the previous case, the operator $\constr_{q,t}$ is triangular and it has a 1-dimensional kernel, spanned by the solution $\Z^{N_f = 2}(\tau)$ of the recursion. Again, $\constr_{q,t}$ can be split as \eqref{eq:D-W} into a diagonal part
\begin{equation}
 D_{q,t} = t^{1-N} q A_{2}
 \sum_{n=0}^\infty \tau_{n}
 \frac{1-t^{n}}{1-t}q^{-n}\partial_{n}
\end{equation}
and an off-diagonal part which we collectively call $W_{q,t}$. While $D_{q,t}$ has degree zero, $W_{q,t}$ is not homogeneous and it contains terms of degree 0, 1 and 2. The recursion in this case has a longer but still finite step and an exact solution can be computed from the initial condition $c_{\emptyset}$ by solving the equation
\begin{equation}
 D_{q,t}\Z^{N_f=2}(\tau) = W_{q,t}\Z^{N_f=2}(\tau)
\end{equation}
order by order in the times $\{\tau_s\}$.
For concreteness we present the solution in terms of correlators up to degree 3,
\begin{equation}
\label{eq:correlators_nf2}
\begin{aligned}
c_{\{3\}} & = -\frac{A_1 \left(t^{{N}}-1\right) }{A_2^3 t^2 \left(t^3-1\right)} \times\\
& \times \Bigg(A_2 \Big(\left(q^2+q+1\right) t^{2 {N}}+\left(q^2+q+1\right) t^{2 {N}+1}+\left(q^2+q-2\right) t^{2 {N}+2}-3 t^{{N}+2}-3 t^2\Big)+\\
&+A_1^2 t^2 \left(t^{{N}}+t^{2 {N}}+1\right)\Bigg) c_{\emptyset}, \\
 c_{\{2,1\}} & = -\frac{A_1 \left(t^{{N}}-1\right) }{A_2^3 (t-1)^2 t^2 (t+1)} \times \\
 & \times \Bigg(A_2 \Big(-\left(q^2-1\right) t^{2 {N}}+\left(q^2+q-2\right) t^{2 {N}+2}-(q+1) t^{{N}+1}-(q+1) t^{{N}+2}+(q+1) t^{2 {N}+1}+2 t^2\Big)+\\
 & +A_1^2 t^2 \left(t^{2 {N}}-1\right)\Bigg)c_{\emptyset}, \\
 c_{\{1,1,1\}} & = -\frac{A_1 \left(t^{{N}}-1\right) \left(A_2 (q-1) (t-1) t^{{N}} \left(-(q-1) t^{{N}}+(q+2) t^{{N}+1}-3 t\right)+A_1^2 t^2 \left(t^{{N}}-1\right)^2\right)}{A_2^3 (t-1)^3 t^2} c_{\emptyset}, \\
 c_{\{2\}} & = \frac{\left(t^{{N}}-1\right) \left(A_2 \left((q+1) t^{{N}}+(q-1) t^{{N}+1}-2 t\right)+A_1^2 t \left(t^{{N}}+1\right)\right)}{A_2^2 t \left(t^2-1\right)} c_{\emptyset}, \\
 c_{\{1,1\}} & = \frac{\left(t^{{N}}-1\right) \left(A_2 (q-1) (t-1) t^{{N}}+A_1^2 t \left(t^{{N}}-1\right)\right)}{A_2^2 (t-1)^2 t} c_{\emptyset}, \\
 c_{\{1\}} & = -\frac{A_1 \left(t^{{N}}-1\right)}{A_2 (t-1)} c_{\emptyset}~,
\end{aligned}
\end{equation}
which is consistent with the result in \cite{Cassia:2019sjk} (if we identify $A\equiv A_1$ and $B\equiv A_2$). We observe that the correlators are rational functions of the parameters $q$, $t$, $A_{1,2}$, and that in the large $N$ limit they behave as
\begin{equation}
 c_{\rho}/c_{\emptyset} \sim t^{N|\rho|}~.
\end{equation}

\subsubsection*{Semi-classical limit}

In order to make the semi-classical limit $q=\mathe^{\hbar}$ with $\hbar\to0$ well-defined at the level of the shift of times in \eqref{eq:shiftoftimesD2S1other}, we need to appropriately choose the masses $u_1$, $u_2$ so that they scale non-trivially with $\hbar$ in such a way that the shift is finite. By imposing the conditions
\begin{equation}
 p_{1}(u) = q^{-1}(1-q)a_1,
 \quad\quad\quad
 p_{2}(u) = q^{-2}(1-q^2)a_2
\end{equation}
for $a_1$, $a_2$ two (positive) constants, we can parametrize the choice of masses as
\begin{equation}
\label{eq:masseslimitNf2}
\begin{aligned}
 u_1 & = \frac{(1-q)a_1+\sqrt{(1-q)\left(a_1^2(q-1)+2a_2(q+1)\right)}}{2q} \\
 u_2 & = \frac{(1-q)a_1-\sqrt{(1-q)\left(a_1^2(q-1)+2a_2(q+1)\right)}}{2q}
\end{aligned}
\end{equation}
up to permutation $u_1\leftrightarrow u_2$.
The anti-chiral 1-loop determinant of the fundamental anti-chirals becomes
\begin{equation}
 \prod_{i=1}^N \prod_{k=1,2} (q\lambda_i u_k;q)_\infty
 = \exp \left(-a_1\sum_{i=1}^N\lambda_i-\frac{a_2}{2}\sum_{i=1}^N\lambda_i^2 + \dots \right)
\end{equation}
where the coefficients of the higher powers of $\lambda_i$ are completely determined by $a_{1,2}$ and, more importantly, they all vanish in the limit $q\to1$.
With this parametrization we also have that the variables $A_1$, $A_2$ behave as
\begin{equation}
 A_{k} = a_{k} \hbar + O(\hbar^2)~.
\end{equation}
Expanding the operator $\constr_{n}$ in powers of $\hbar$ as in \eqref{eq:hbarexpansionZ}, the first non-trivial contribution is
\begin{equation} \label{eq:classicalVirasoroNf2}
\begin{split}
 -\constr^{(1)}_n = & 2 \beta N \partial_n + 
 \beta \sum_{a+b=n} \partial_a\partial_b + (1-\beta)(n+1)\partial_n + \delta_{n,-1} \tau_1 N + \delta_{n,0} \left( \beta N^2 + (1-\beta)N \right) \\
 & +\sum_{s>0} s \tau_s \partial_{s+n}-\delta_{n,-1}a_1 N - a_1\partial_{n+1} - a_2\partial_{n+2}
\end{split}
\end{equation}
which we immediately recognize as the Virasoro constraint operator for the classical matrix model
in \eqref{eq:p2classicalgenfun}.
By re-summing with weight $(n+2)\tau_{n+2}$ over $n\geq-1$, we get the relation
\begin{equation}
 -\constr_{q,t}^{(1)} =  W_{-2} - {a_1} L_{-1} - a_2 D  
\end{equation}
which matches exactly with the classical constraint \eqref{eq:p2classicalconstr}. We conclude that the generating functions $\Z^{N_f=2}_{(0)}(\tau)$ and $\Z^{\p=2}(\tau)$ are equivalent as they are both defined as the generators of the 1-dimensional vector space $\ker\constr_{q,t}^{(1)}$.

\subsubsection*{Averages of Macdonald Polynomials}

Evaluating the average of Macdonald polynomials on the explicit solution that we found, we are able to write the following identity
\begin{equation}
\label{eq:macdonaldNf2}
 \lbracket\macdonald_\rho(p_k)\rbracket^{N_f=2} =
 \frac{\macdonald_\rho\left(
 p_k=(-1)^k t^{\frac{k}{2}{N}} \frac{\left(u_1^{k}+u_2^{k}\right)}{1-t^k}
 \right)}
 {\macdonald_\rho\left(
 p_k=(-1)^k t^{\frac{k}{2}} \frac{ (u_1 u_2)^{k}}{1-t^k}
 \right)}
 \macdonald_\rho(p_k=\pi_k^{(N)}) \, c_{\emptyset}~,
\end{equation}
which we checked explicitly for all partitions of degree $6$ and lower. The $\pi_k^{(N)}$ are defined as in \eqref{eq:pikNdef}.
\\
\\
Assuming that this formula holds for all partitions $\rho$, we can compute the semi-classical limit $t=q^\beta$ and $q\to1$ with the choice of masses in \eqref{eq:masseslimitNf2}. The result matches exactly with formula \eqref{eq:jackNf2} for the average of Jack polynomials at $\p=2$.

\subsubsection{Comments on $N_f\geq3$}

Our analysis indicates that the only cases that admit a full and unique solution (independent of the normalization of the empty correlator) are those of $N_f=1,2$. For higher values of $N_f$ we have a similar situation to that of the classical models at $\p\geq3$. The $q$-Virasoro constraints have an infinite dimensional kernel, corresponding to the fact that one needs to specify more initial conditions to solve the recursion.
{We believe that the situation here mirrors what we discussed in Section~\ref{sec:p>3}, however the actual formulas are too cumbersome and unilluminating to write down explicitly. We remark however that computer calculations suggest that the actual recursion relations can be solved iteratively by the same procedure delineated there.
We do however expect that a more subtle study of possible analytical issues is required in this $q$-deformed case, especially with regards to the dependence on the mass parameters $u_k$. The definition of the matrix model for instance might have ambiguities related to the choice of contour, similar to the so called Dijkgraaf-Vafa phase of a classical matrix model \cite{Dijkgraaf:2002fc,Dijkgraaf:2002vw}.}
\\
\\
The semi-classical behavior is also not so straightforward.
This limit can, in fact, be defined by the choice of masses which solves the equations
\begin{equation}
 p_k(u) = q^{-k}(1-q^k)a_k~,
 \quad\quad\quad
 k=1,\dots,N_f~.
\end{equation}
In terms of the $u_k$, these form a system of $N_f$ equations of increasing degree up to $N_f$. Over the complex numbers there are $N_f!$ solutions which are all equivalent upon permutations of the $u_k$, however writing such a solution explicitly is in general not possible. What we expect is that upon substitution in \eqref{eq:shiftoftimesD2S1other} we get a shift of the times such that only the first $N_f$ terms are non-vanishing in the limit $q\to1$. Then the constants $a_k$ parametrizing the solution, can be identified with the coupling constants of the semi-classical model where $\p=N_f$.

\section{Conclusion}
\label{sec:conclusion}

In this paper we give an outline of the procedure to obtain and recursively solve the Virasoro and $q$-Virasoro constraints in the case of the {\it classical} $\beta$-deformed Hermitean 1-matrix model and the {\it quantum}  3d $\mathcal{N}=2$ theory with $U(N)$ gauge group on $\halfindex$ and $S_b^3$. We present the solution of the models in terms of explicit expressions for the first few correlators, and additionally in the classical case we can also express the solution using the $W$-representation of the generating function. Moreover, we deduce novel formulas of the form $\langle \mathsf{character}\rangle = \mathsf{character}$ in the spririt of \cite{Morozov:2018eiq} as given in \eqref{eq:jackNf1}, \eqref{eq:jackNf2}, \eqref{eq:macdonaldNf1} and \eqref{eq:macdonaldNf2}.
Finally, we explicitly match the classical models with their $q$-deformations by showing the existence of a well-defined semi-classical expansion around a formal deformation parameter $q=\mathe^{\hbar}$, for small $\hbar$.
\\
\\
There are several directions for further investigation. 
\begin{itemize}
\item One obvious direction is to extend the $W$-representation of the generating function to the quantum models, an investigation which has been initiated in \cite{Morozov_2019}. This is a much more difficult task than in the classical case due to the more involved constraint equations.
\item Another interesting direction might be to try to find an analytical proof of the formulas for expectation values of characters. These were all unexpected a priori, and the formulas appearing in this paper have only been verified by explicitly checking the equations for partitions up to some finite degree. 
\item Another open question is regarding the condition \eqref{eq:effectiveCS} on the effective CS level. In particular how one can treat models when $\kappa_2^{\textrm{eff}}$ is different from zero. 
\item Furthermore, another question is whether the procedure of obtaining and solving the Virasoro constraints can be generalized to other root systems, in the sense that the above derivation is valid only for a $U(N)$ gauge group or $A_N$ from a matrix model perspective. The question is then if there are other versions of the derivation for other Lie groups, which then both offers a definition of the matrix model, together with a set of Virasoro constraints which may or may not be solvable. 
\item Another interesting question to consider, is why the system of equations in the Virasoro constraints are triangular and give rise to a finite recursion relation. From the explicit Virasoro constraints it is clear that the equations are triangular, but why this is so in the first place is not obvious.
\item {There is also the question if there is a deeper reason to why $\p \geq 3$ in the classical case and $N_f \geq3$ in the quantum case cannot be solved completely. At the level of the constraint equations we understand why this is not possible, but it is still not clear if there is a more fundamental physical reason to why this is the case. Following the parallel with the classical models at $p\geq 3$ one is lead to expect that the quantum models at $N_f\geq 3$ present ambiguities in the $q$-Virasoro solution due to the presence of multiple inequivalent phases of the Dijkgraaf-Vafa type, which would be induced by Stokes phenomena appearing in the choice of integration contour. Further analytical studies are required to determine if this picture is correct.}
\item {Finally, one can speculate on the physical implications of our results for the corresponding 3d $\mathcal{N}=2$ gauge theories. It was observed in \cite{Nazzal:2018brc,Razamat:2018zel} that a certain type of $q$-difference operator acting on a 4d $\mathcal{N}=1$ index could be realized explicitly as the insertion of a supersymmetric surface defect in spacetime. One might reasonably expect that the Macdonald operator in \eqref{eq:qDifferenceOperator} also can be realized similarly as the insertion of a supersymmetric defect in 3d. In particular, since the 4d surface defect wraps around the $S^1$ direction, upon compactification to 3d one should expect the defect to become a line defect wrapping some equator of the $S^3_b$. We leave this question for future investigations.}
\end{itemize}

\section*{Acknowledgments}
We thank Anton Nedelin and Aleksandr Popolitov for useful discussions.
All authors are supported in part by the grant ``Geometry and Physics'' from the Knut and Alice Wallenberg foundation.

\appendix

\section{Special functions}
\label{sec:special_functions}

Here we recall the special functions that we use throughout the paper. Firstly, the $q$-Pochhammer symbol is defined as
\begin{equation}
\label{eq:qpochhammer}
 (z;q)_\infty = \prod_{k=0}^\infty (1-z q^k)~,
\end{equation}
with $z\in\mathbb{C}$ and $|q|<1$. The analytic continuation to the region $|q|>1$ is given by 
\begin{equation}
\label{eq:qpochhammer_continued}
 (z;q)_\infty = \frac{1}{\left( q^{-1}z;q^{-1}\right)_\infty}~.
\end{equation}
When $|z|<1$ and $|q|\neq1$ one can define the quantum dilogarithm \cite{Faddeev:1993rs}
\begin{equation}
\label{eq:qdilog}
 \mathrm{Li}_2(z;q) = \sum_{n=1}^\infty \frac{z^n}{n(1-q^n)}~,
\end{equation}
which can be used to rewrite the $q$-Pochhammer as
\begin{equation}
\label{eq:qpochhammeridentity}
 (z;q)_\infty = \exp\left(-\mathrm{Li}_2(z;q)\right)~.
\end{equation}
Secondly, we introduce the double sine function \cite{Kurokawa:2003,Narukawa:2003}.
For $\underline{\omega}\equiv(\omega_1,\omega_2)\in\mathbb{C}^2$ with $\mathrm{Re}(\omega_1)>0$, $\mathrm{Re}(\omega_2)>0$ and $z\in\mathbb{C}$,
the double sine function is defined by the regularized infinite product
\begin{equation}
\label{eq:doubleSine}
  S_2 \left(z|\underline{\omega}\right) = \prod_{n_1, n_2
  \geq 0} \frac{n_1 \omega_1 + n_2 \omega_2 + z}{n_1 \omega_1 + n_2 \omega_2 + \omega
  - z}~, \quad\quad\quad \omega = \omega_1 + \omega_2~,
\end{equation}
which satisfies the inversion property
\begin{equation}
\label{eq:doubleSineInversion}
  S_2 \left(z|\underline{\omega}\right)S_2\left(\omega-z|\underline{\omega}\right)=1~.
\end{equation}
For generic values of the parameters $\omega_1,\omega_2$ such that $\mathrm{Im}(\frac{\omega_2}{\omega_1})\neq 0$, the double Sine function has the following infinite product representation
\begin{equation}
\label{eq:S2identity}
 S_2(z|\underline{\omega})=\mathe^{\frac{\mathi\pi}{2}B_{22}(z|\underline{\omega})}\left(\mathe^{\frac{2\pi\mathi}{\omega_1}z};\mathe^{2\pi\mathi\frac{\omega}{\omega_1}}\right)_\infty \left(\mathe^{\frac{2\pi\mathi}{\omega_2}z};\mathe^{2\pi\mathi\frac{\omega}{\omega_2}}\right)_\infty~,
\end{equation}
where $B_{22}(z|\underline\omega)$ is the double Bernoulli polynomial
\begin{equation}
\label{eq:B22}
B_{22}(z|\underline\omega)=\frac{1}{\omega_1\omega_2}\left(\left(z-\frac{\omega}{2}\right)^2-\frac{\omega_1^2+\omega_2^2}{12}\right)~.
\end{equation}

\section{Symmetric functions and characters}
\label{sec:characters}

The Schur polynomials denoted by $\schur_{\gamma}(\lambda_1,\dots,\lambda_N)$ are labeled by partitions $\gamma$ and are defined as the irreducible characters of the group $U(N)$. As such they form an orthonormal (linear) basis in the space of all polynomial characters which, by definition, are invariant under the action of the Weyl group $S_N$. This means that Schur polynomials form a basis for all symmetric functions.
\\
\\
For any partition $\gamma=\{\gamma_1,\dots,\gamma_N\}$ whose elements obey $\gamma_1\geq\dots\geq\gamma_N\geq0$ one can compute the Schur polynomial $\schur_{\gamma}(\lambda_1,\dots,\lambda_N)$ as a ratio of determinants as in the Weyl character formula
\begin{equation}
\label{eq:schurgeneral}
 \schur_{\gamma}(\lambda_1,\dots,\lambda_N) = \frac{\det\limits_{i j} \, \lambda_{i}^{N+\gamma_j-j}}{\det\limits_{i j} \, \lambda_{i}^{N-j}}~.
\end{equation}
Introducing the power-sum variables $p_k=\sum_{i=1}^N \lambda_i^k$ one can expand Schur polynomials as
\begin{equation}
\label{eq:powertoschur}
 \schur_{\gamma}(p_k) = \sum_{\rho} \frac{\chi^\gamma_\rho}{|{\Aut}(\rho)|}\prod_{a\in\rho}\frac{p_a}{a}~,
\end{equation}
where $\chi^\gamma_\rho$ is the character of the representation of the symmetric group indexed by the partition $\gamma$ evaluated at elements of cycle type $\rho$. Here we also introduced the notation ${\Aut}(\rho)$ for the automorphism group of the partition $\rho$, namely the group of permutations of the parts of $\rho$ which are of equal length. The order of this group can be computed as
\begin{equation}
 |{\Aut}(\rho)| = \prod_{a\in\rho}\frac{\partial}{\partial\tau_a}
 \cdot \prod_{b\in\rho}\tau_b~.
\end{equation}
For symmetric Schur polynomials \eqref{eq:powertoschur} takes the simpler form
\begin{equation}
\label{eq:symmschur}
 \schur_{\{m\}}(p_k) =
 \sum_{\gamma\vdash m}
 \frac{1}{|\Aut(\gamma)|}
 \prod_{a\in\gamma} \frac{p_a}{a}~,
\end{equation}
where $\gamma\vdash m$ denotes that $\gamma$ is an integer partition of $m$.
\\
\\
If we define the scalar product of two symmetric functions by
\begin{equation}
\label{eq:hallinnerprod}
 (f|g) = \frac{1}{(2\pi\mathi)^N N!}\oint_{|\lambda|=1}\prod_{i=1}^N\frac{\mathd\lambda_i}{\lambda_i}
 \Delta(\lambda)\Delta(\lambda^{-1}) f(\lambda)g(\lambda^{-1})~,
\end{equation}
then Schur polynomials are orthogonal
\begin{equation}
 (\schur_\gamma|\schur_\rho) = \delta_{\gamma,\rho}~.
\end{equation}
Schur polynomials also satisfy the Cauchy identity \cite[Chapter I, (4.3)]{Macdonald}
\begin{equation}
\label{eq:cauchy}
 \exp \left( \sum_{k=1}^\infty \frac{\tau_k p_k}{k} \right) = \sum_\gamma \schur_\gamma\left(  \tau_k\right) \schur_\gamma\left( p_k \right)~,
\end{equation}
where the summation on the right hand side is over all partitions $\gamma$.
Using the plethystic substitution given by $\tau_k=z^k$ together with \eqref{eq:symmschur} we obtain the useful formula
\begin{equation}
\label{eq:symm-schur-poly}
 \exp\left(\sum_{k=1}^\infty \frac{z^k p_k}{k}\right) =
 \sum_{m=0}^\infty z^m \schur_{\{m\}}(p_1,\dots,p_m)~.
\end{equation}
The Jack and Macdonald polynomials are defined as 1- or 2-parameter deformations of the Schur polynomials (see \cite{Macdonald}). More specifically, for any $\beta\in\mathbb{R}_{>0}$ we define Jack polynomials $\jack_\gamma (p_k)$ as the symmetric functions orthogonal with respect to the inner product
\begin{equation}
\label{eq:hallinnerprodJack}
 (f|g)_\beta = \frac{1}{(2\pi\mathi)^N N!}\oint_{|\lambda|=1}
 \prod_{i=1}^N\frac{\mathd\lambda_i}{\lambda_i}
 \left(\Delta(\lambda)\Delta(\lambda^{-1})\right)^\beta f(\lambda)g(\lambda^{-1})~.
\end{equation}
Given parameters $|q|<1$ and $|t|<1$ we can define Macdonald polynomials $\macdonald_\gamma(p_k)$ as a family of symmetric functions orthogonal with respect to the inner product
\begin{equation}
\label{eq:hallinnerprodMacdonald}
 (f|g)_{q,t} = \frac{1}{(2\pi\mathi)^N N!}\oint_{|\lambda|=1}
 \prod_{i=1}^N\frac{\mathd\lambda_i}{\lambda_i}
 \Delta_{q,t}(\lambda) f(\lambda)g(\lambda^{-1})~.
\end{equation}
For concreteness we only consider Jack and Macdonald polynomials in the $P$-basis.
In the limit $t\to q$, Macdonald polynomials degenerate to Schur polynomials, while for $t=q^\beta$ they give the Jack polynomials.

\section{Relating constraint generators to $q$-Virasoro generators}\label{sec:generators_relation}

We now wish to obtain a relation between the generators of the constraint $\constr_n$ in the quantum case and the generators of the $q$-Virasoro algebra $\hat{T}_n$. Similar to the analysis in \cite{Cassia:2019sjk}, one can  introduce the function $\psi(z)$ defined as
\begin{equation}\label{eq:psi}
\psi(z) = q^{-1/2} t^{1/2} r^{-1/2} \exp\left( \sum_{s=1}^\infty z^{-s} \frac{(1-q^s)}{(1+q^st^{-s})}\tau_s \right) 
\end{equation}
and the generator current $\hat{T}(z) = \sum_{n \in \mathbb{Z}} \hat{T}_n z^n$ as
\begin{equation}
\begin{split}
\hat{T}(z) = & q^{1/2}t^{-1/2}r^{1/2}
 \exp\left(-\sum_{s=1}^\infty z^{-s}\frac{(1-q^s)}{(1+q^st^{-s})}\tau_s \right) 
 \exp\left(\sum_{s=1}^\infty z^{s}\frac{(1-t^{-s})}{s}\partial_{s}\right)t^{N} + \\
 & + q^{1/2} t^{1/2} r^{-1/2}
 \exp\left( \sum_{s=1}^\infty z^{-s}(1-q^s)\left(\frac{\tau_{s} q^st^{-s}}{(1+q^st^{-s})}
 +\frac{p_{s}(u)}{s(1-q^{-s})}\right) \right)
 \exp\left(\sum_{s=1}^\infty z^{s} \frac{(1-t^s)}{s q^s}
 \partial_{s} \right)t^{-N}~.
\end{split} 
\end{equation}
Here the only differences compared to the standard generator current $T(z)$ of the $q$-Virasoro algebra \cite{Shiraishi:1995rp}, 
\begin{equation}
\begin{split}
T(z) = & q^{1/2}t^{-1/2} \exp\left(- \sum_{s=1}^\infty z^{-s} \frac{(1-q^s)}{(1+q^st^{-s})}\tau_s \right) 
 \exp\left(\sum_{s=1}^\infty z^{s}\frac{(1-t^{-s})}{s}\partial_{s}\right)t^{N} + \\
& \qquad + q^{1/2} t^{1/2}
 \exp\left( \sum_{s=1}^\infty z^{-s}\frac{(1-q^s) q^st^{-s}}{(1+q^st^{-s})}\tau_{s}
  \right)
 \exp\left(\sum_{s=1}^\infty z^{s} \frac{(1-t^s)}{s q^s}
 \partial_{s} \right)t^{-N}~,
\end{split} 
\end{equation}
are therefore the factors of $r^{\pm 1/2}$ and the shift of times proportional to $p_s(u)$.
\\
\\
We can also denote the eigenvalues of the generator $\hat{T}_n$ for $n \geq -1$ as
\begin{equation}
\hat{T}_n \Z_{\halfindex}^{N_f} (\tau) = \eval_n \, \Z_{\halfindex}^{N_f} (\tau)~. 
\end{equation}
That the operator $\hat{T}_n$ is diagonal also for $n=-1$ is not obvious, although we show below that this is the case.
\\
\\
Then, we can rewrite the left hand side of the constraints in \eqref{eq:qVirasoroD2S1} as
\begin{equation}\label{eq:lhs_constraint}
\frac{1}{1-t} \, \psi(z)\hat{T}(z) \Z_{\halfindex}^{N_f}(\tau) 
\end{equation}
so that one can read off the eigenvalue for the $\hat{T}_0$ by equating \eqref{eq:lhs_constraint} and the right hand side of \eqref{eq:qVirasoroD2S1} to find
\begin{equation}
\hat{T}_0 \Z_{\halfindex}^{N_f} (\tau) = \left( q^{1/2} t^{-1/2} r^{1/2} +q^{-1/2} t^{1/2} r^{-1/2} \right) \Z_{\halfindex}^{N_f}(\tau) ~.
\end{equation}
For the $\hat{T}_{-1}$ operator, the eigenvalue is only well defined in the case of $\dvar \rightarrow 0$ or $r=q^{\dvar} \rightarrow 1 $ (when we also recover equation (3.37) of \cite{Cassia:2019sjk}), in which case it is
\begin{equation}
\hat{T}_{-1} \Z_{\halfindex}^{N_f}(\tau) = -q^{1/2}t^{1/2}p_1(u) \Z_{\halfindex}^{N_f}(\tau) ~.
\end{equation}
From the dependence on $p_1(u)$ in the above, we can also see that the eigenvalue of $T_{-1}$ vanishes.
\\
\\
We now recall that the $q$-Virasoro constraints were written using the generators $\constr_n$ as $\constr_{n \geq 0} \Z_{\halfindex}^{N_f}(\tau)=0$.
Then the connection between the generators $\constr_s$ and the generator current $\hat{T}(z)$ can be written as 
\begin{equation}
\constr_s \Z_{\halfindex}^{N_f} (\tau) = \oint_{z=0} \frac{\mathd z}{2 \pi \mathi z} z^{-s} \psi (z)\left( \hat{T} (z) - \frac{\eval_{-1}}{z}- \eval_0 \right)  \Z_{\halfindex}^{N_f} (\tau)
\end{equation}
valid for $s=-1,0, \dots $, where taking the residue is simply a way of extracting the coefficient of $z^s$ in the $z$-expansion.

\section{Asymptotic analysis and convergence of $S^3_b$ partition function}\label{sec:asymptotic_analysis}

\begin{figure}[ht]
  \caption{Closed integration contour for evaluating the partition function $\Z_{S^3_b}$.
  The crosses indicate the schematic locations of simple poles.}
  \label{fig:residue}
	\begin{center}
    \resizebox{0.4\textwidth}{!}{%
	\begin{tikzpicture}[x=0.5cm,y=0.5cm]
	\begin{scope}
	\clip (0,0) circle (5);
	\end{scope}
	\draw [->,gray](-6.5,0)--(6.5,0) node [right, black] {$\mathbb{R}$};
	\draw [->,gray](0,-5.5)--(0,5.5);
	\node at (-1.5,5.3) [right,black] {$\mathi\mathbb{R}$};
	\draw [->,very thick,black](0,-5)--(0,5) arc(90:-90:5);
	\draw [thin,gray](5,5)--(5,4)--(6,4);
	\node at (5.65,4.6) {$X_i$};
	\node at (0.7,0) {\small $\times$};	
	\node at (1.4,0) {\small $\times$};
	\node at (2.1,0) {\small $\times$};
	\node at (3.5,0) {\small $\times$};
	\node at (4.2,0) {\small $\times$};	
	\end{tikzpicture}
    }
	\end{center}
\end{figure}
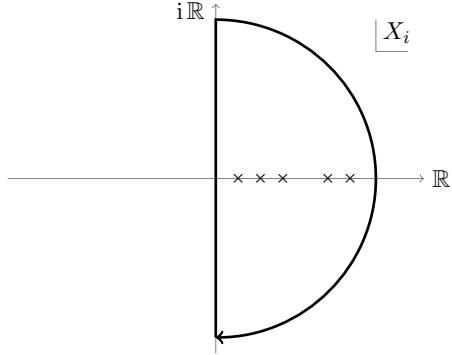

In this section we address possible analytical issues in the definition of the partition function on the squashed 3-sphere. In particular, we study the asymptotic behavior of the integrand to determine the conditions under which the partition function and the correlators are convergent.
\\
\\
From \cite[Theorem 2.2]{Rains:2006dfy} we have that for $\omega_1,\omega_2\in\mathbb{R}_{>0}$ the function
\begin{equation}
\label{eq:bounded}
 S_2(z|\underline{\omega}) \exp\left(-\epsilon\frac{\pi\mathi(z-\omega/2)^2}{2\omega_1\omega_2}
 \right)
 \quad\quad\quad
 \text{with }\epsilon:=\mathrm{sign}(\mathrm{Im}(z))
\end{equation}
is bounded at infinity (see also \cite[Proposition 2.2.6 and Corollary 5.2.7]{vdBult:2007}).
\\
\\
Let us look at the fundamental anti-chiral contribution to the integrand. These are of the form
\begin{equation}
 S_2(-X_i-m_k|\underline{\omega})^{-1} = S_2(X_i+m_k+\omega|\underline{\omega})
\end{equation}
where we used the inversion formula \eqref{eq:doubleSineInversion}. Plugging $z=X_i+m_k+\omega$ in \eqref{eq:bounded} we obtain the asymptotic approximation
\begin{equation}
 \label{eq:asymp2}
 S_2(-X_i-m_k|\underline{\omega})^{-1} \approx
 \exp\left(-\frac{\pi\mathi}{2\omega_1\omega_2}(-\epsilon)X_i^2 + \frac{2\pi\mathi}{\omega_1\omega_2}\epsilon\left(\frac{m_k}{2}+\frac{1}{2}\frac{\omega}{2}\right) X_i \right)~.
\end{equation}
Therefore, at large $|X_i|$ the fundamental chirals behave as shifts in the CS level and FI parameter.
Putting together all the potential terms we get
\begin{equation}
\begin{aligned}
 \underbrace{\mathe^{-\frac{\pi\mathi\kappa_2}{\omega_1\omega_2} X_i^2}}_{\mathrm{CS}}
 \underbrace{\mathe^{\frac{2\pi\mathi\kappa_1}{\omega_1\omega_2} X_i}}_{\mathrm{FI}} &
 \prod_{k=1}^{N_f}\underbrace{S_2(-X_i-m_k|\underline{\omega})^{-1}}_{\text{anti-chirals}}
 \approx \\
 & \approx\exp\left\{-\frac{\pi\mathi}{2\omega_1\omega_2}\left(2\kappa_2 -\epsilon N_f\right)X_i^2
 +\frac{2\pi\mathi}{\omega_1\omega_2}
  \left(\kappa_1+\epsilon\sum_{k=1}^{N_f}\frac{m_k}{2}+\epsilon\frac{N_f}{2}\frac{\omega}{2}\right)
 X_i\right\}~.
\end{aligned}
\end{equation}
Similarly, the measure $\Delta_S(X)$ introduces a shift of the FI parameter coming from the contribution\footnote{In order to see this, one should write $\Delta_S(X)$ as the product $\prod_{\alpha=1,2}\Delta_{q,t}(\lambda_{\alpha})$ and then apply \eqref{eq:qtmeasurerewrite}.}
\begin{equation}
 \prod_{\alpha=1,2}\lambda_{i,\alpha}^{-\beta(N-1)}
 = \exp\left(\frac{2\pi\mathi}{\omega_1\omega_2}(-M_\mathrm{a}(N-1))X_i\right)
\end{equation}
so that the integrand asymptotically behaves as the function
\begin{equation}
\exp\left\{-\frac{\pi\mathi}{2\omega_1\omega_2}\left(2\kappa_2 -\epsilon N_f\right)X_i^2
 +\frac{2\pi\mathi}{\omega_1\omega_2}
  \left(\kappa_1+\epsilon\sum_{k=1}^{N_f}\frac{m_k}{2}+\epsilon\frac{N_f}{2}\frac{\omega}{2}
 -M_\mathrm{a}(N-1)\right)
 X_i\right\}~.
\end{equation}
Finally, we have:
\begin{itemize}
 \item For $\mathrm{Im}(X_i)<0$ (i.e. $\epsilon=-1$), $2\kappa_2+N_f>0$ so that the quadratic term is dominant. The exponential goes to zero at infinity if $\mathrm{Re}(X_i)>0$.
 \item For $\mathrm{Im}(X_i)>0$ (i.e. $\epsilon=+1$), $2\kappa_2-N_f=0$ and the quadratic term $X_i^2$ vanishes. The linear term becomes dominant and the exponential goes to zero at infinity if
\begin{equation}
\label{eq:convergence}
 \mathrm{Re}\left(\omega(\dvar+1)\right) = \mathrm{Re}\left(\kappa_1+\sum_{k=1}^{N_f}\frac{m_k}{2}+\frac{N_f}{2}\frac{\omega}{2}-M_\mathrm{a}(N-1)\right)
  > 0~.
\end{equation}
In our conventions $\omega\in\mathbb{R}_{>0}$, so that this constraint is equivalent to the requirement $\mathrm{Re}(\dvar)>-1$ which corresponds to the condition for convergence of the integral that we observed in the classical model in \eqref{eq:WishartLaguerre}. For $N_f>1$ we have $\dvar=0$ so that \eqref{eq:convergence} is trivially satisfied.
\end{itemize}
This analysis then implies that the integral can be computed by closing the contour in the right-hand-plane and taking the residue as in Figure~\ref{fig:residue}.
\\
\\
Physically speaking, one can map the real part of the complex masses to the R-charge of the corresponding field as follows: $\mathrm{Re}(m_k)=-\frac{\omega}{2}R_k$ and
$\mathrm{Re}(M_\mathrm{a})=\frac{\omega}{2}R_\mathrm{a}$, with $R_k$ the R-charge of the $k$-th fundamental field and similarly $R_\mathrm{a}$ is the R-charge of the adjoint field, while $\mathrm{Re}(\kappa_1) = \frac{\omega}{2}R_{\mathrm{monopole}}$ is the R-charge of the bare monopole operator \cite{Willett:2011gp}. Rewriting \eqref{eq:convergence} in terms of the R-charges of the fermions and using the fact that the gauginos carry R-charge 1, we have
\begin{equation}
 R_{\mathrm{monopole}}^\mathrm{eff} = R_{\mathrm{monopole}}
 -\frac{1}{2}\sum_{k=1}^{N_f}(R_k-1) - (R_\mathrm{a}-1)(N-1)-(N-1)
 = 2\left(\mathrm{Re}(\dvar)+1\right) > 0~,
\end{equation}
so that convergence of the integral imposes the positivity of the effective R-charge of the monopole operator, while the balancing condition ($\dvar=0$) further restricts this charge to be 2.
This considerations suggest that there might be non-perturbative effects giving rise to a monopole superpotential when $\dvar=0$ similar to those studied in \cite{Amariti:2018wht}.

\providecommand{\href}[2]{#2}\begingroup\raggedright\endgroup


\begin{thebibliography}{10}

\bibitem{Nekrasov-BPS/CFT-1}
N.~Nekrasov, ``{On the BPS/CFT correspondence},'' {\em Lecture at the
  University of Amsterdam string theory group seminar (Feb. 3, 2004)} .

\bibitem{Nekrasov-BPS/CFT-2}
N.~Nekrasov, ``{2d CFT-type equations from 4d gauge theory},'' {\em Lecture at
  the IAS conference -Langlands Program and Physics (March 8-10, 2004)} .

\bibitem{Pestun:2016zxk}
V.~Pestun {\em et~al.}, ``{Localization techniques in quantum field
  theories},'' \href{http://dx.doi.org/10.1088/1751-8121/aa63c1}{{\em J. Phys.}
  {\bfseries A50} no.~44, (2017) 440301},
\href{http://arxiv.org/abs/1608.02952}{{\ttfamily arXiv:1608.02952 [hep-th]}}.

\bibitem{Mironov:1990im}
A.~Mironov and A.~Morozov, ``{On the origin of Virasoro constraints in matrix
  models: Lagrangian approach},''
\href{http://dx.doi.org/10.1016/0370-2693(90)91078-P}{{\em Phys. Lett.}
  {\bfseries B252} (1990) 47--52}.

\bibitem{Dijkgraaf:1990rs}
R.~Dijkgraaf, H.~L. Verlinde, and E.~P. Verlinde, ``{Loop equations and
  Virasoro constraints in nonperturbative 2-D quantum gravity},''
  \href{http://dx.doi.org/10.1016/0550-3213(91)90199-8}{{\em Nucl. Phys.}
  {\bfseries B348} (1991) 435--456}.
[,435(1990)].

\bibitem{Morozov:1994hh}
A.~Morozov, ``{Integrability and matrix models},''
  \href{http://dx.doi.org/10.1070/PU1994v037n01ABEH000001}{{\em Phys. Usp.}
  {\bfseries 37} (1994) 1--55},
\href{http://arxiv.org/abs/hep-th/9303139}{{\ttfamily arXiv:hep-th/9303139
  [hep-th]}}.

\bibitem{Odake:1999un}
S.~Odake, ``{Beyond CFT: Deformed Virasoro and elliptic algebras},'' in {\em
  Theoretical physics at the end of the twentieth century. Proceedings, Summer
  School, Banff, Canada, June 27-July 10, 1999}, pp.~307--449.
\newblock 1999.
\newblock
\href{http://arxiv.org/abs/hep-th/9910226}{{\ttfamily arXiv:hep-th/9910226
  [hep-th]}}.
\newblock

\bibitem{Beem:2012mb}
C.~Beem, T.~Dimofte, and S.~Pasquetti, ``{Holomorphic Blocks in Three
  Dimensions},'' \href{http://dx.doi.org/10.1007/JHEP12(2014)177}{{\em JHEP}
  {\bfseries 12} (2014) 177},
\href{http://arxiv.org/abs/1211.1986}{{\ttfamily arXiv:1211.1986 [hep-th]}}.

\bibitem{Yoshida:2014ssa}
Y.~Yoshida and K.~Sugiyama, ``{Localization of 3d $\mathcal{N}=2$
  Supersymmetric Theories on $S^1 \times D^2$},''
  \href{http://arxiv.org/abs/1409.6713}{{\ttfamily arXiv:1409.6713 [hep-th]}}.

\bibitem{Imamura:2011wg}
Y.~Imamura and D.~Yokoyama, ``{N=2 supersymmetric theories on squashed
  three-sphere},'' \href{http://dx.doi.org/10.1103/PhysRevD.85.025015}{{\em
  Phys. Rev. D} {\bfseries 85} (2012) 025015},
  \href{http://arxiv.org/abs/1109.4734}{{\ttfamily arXiv:1109.4734 [hep-th]}}.

\bibitem{Hama:2011ea}
N.~Hama, K.~Hosomichi, and S.~Lee, ``{SUSY Gauge Theories on Squashed
  Three-Spheres},'' \href{http://dx.doi.org/10.1007/JHEP05(2011)014}{{\em JHEP}
  {\bfseries 05} (2011) 014}, \href{http://arxiv.org/abs/1102.4716}{{\ttfamily
  arXiv:1102.4716 [hep-th]}}.

\bibitem{Lodin:2018lbz}
R.~Lodin, A.~Popolitov, S.~Shakirov, and M.~Zabzine, ``{Solving $q$-Virasoro
  constraints},'' \href{http://dx.doi.org/10.1007/s11005-019-01216-5}{{\em
  Lett. Math. Phys.} {\bfseries 110} no.~1, (2020) 179--210},
  \href{http://arxiv.org/abs/1810.00761}{{\ttfamily arXiv:1810.00761
  [hep-th]}}.

\bibitem{Nedelin:2016gwu}
A.~Nedelin, F.~Nieri, and M.~Zabzine, ``{$q$-Virasoro modular double and 3d
  partition functions},''
  \href{http://dx.doi.org/10.1007/s00220-017-2882-1}{{\em Commun. Math. Phys.}
  {\bfseries 353} no.~3, (2017) 1059--1102},
\href{http://arxiv.org/abs/1605.07029}{{\ttfamily arXiv:1605.07029 [hep-th]}}.

\bibitem{Morozov:2009xk}
A.~Morozov and S.~Shakirov, ``{Generation of Matrix Models by W-operators},''
  \href{http://dx.doi.org/10.1088/1126-6708/2009/04/064}{{\em JHEP} {\bfseries
  04} (2009) 064},
\href{http://arxiv.org/abs/0902.2627}{{\ttfamily arXiv:0902.2627 [hep-th]}}.

\bibitem{Itoyama:2017xid}
H.~Itoyama, A.~Mironov, and A.~Morozov, ``{Ward identities and combinatorics of
  rainbow tensor models},''
  \href{http://dx.doi.org/10.1007/JHEP06(2017)115}{{\em JHEP} {\bfseries 06}
  (2017) 115}, \href{http://arxiv.org/abs/1704.08648}{{\ttfamily
  arXiv:1704.08648 [hep-th]}}.

\bibitem{Mironov:2017och}
A.~Mironov and A.~Morozov, ``{On the complete perturbative solution of
  one-matrix models},''
  \href{http://dx.doi.org/10.1016/j.physletb.2017.05.094}{{\em Phys. Lett. B}
  {\bfseries 771} (2017) 503--507},
  \href{http://arxiv.org/abs/1705.00976}{{\ttfamily arXiv:1705.00976
  [hep-th]}}.

\bibitem{Cassia:2019sjk}
L.~Cassia, R.~Lodin, A.~Popolitov, and M.~Zabzine, ``{Exact SUSY Wilson loops
  on S$^{3}$ from $q$-Virasoro constraints},''
  \href{http://dx.doi.org/10.1007/JHEP12(2019)121}{{\em JHEP} {\bfseries 12}
  (2019) 121}, \href{http://arxiv.org/abs/1909.10352}{{\ttfamily
  arXiv:1909.10352 [hep-th]}}.

\bibitem{Mironov:2018ekq}
A.~Mironov and A.~Morozov, ``{Sum rules for characters from
  character-preservation property of matrix models},''
  \href{http://dx.doi.org/10.1007/JHEP08(2018)163}{{\em JHEP} {\bfseries 08}
  (2018) 163}, \href{http://arxiv.org/abs/1807.02409}{{\ttfamily
  arXiv:1807.02409 [hep-th]}}.

\bibitem{Morozov:2018eiq}
A.~Morozov, A.~Popolitov, and S.~Shakirov, ``{On $(q,t)$-deformation of
  Gaussian matrix model},''
  \href{http://dx.doi.org/10.1016/j.physletb.2018.08.006}{{\em Phys. Lett.}
  {\bfseries B784} (2018) 342--344},
\href{http://arxiv.org/abs/1803.11401}{{\ttfamily arXiv:1803.11401 [hep-th]}}.

\bibitem{Marino:2012zq}
M.~Mariño, ``{Lectures on non-perturbative effects in large $N$ gauge
  theories, matrix models and strings},''
  \href{http://dx.doi.org/10.1002/prop.201400005}{{\em Fortsch. Phys.}
  {\bfseries 62} (2014) 455--540},
  \href{http://arxiv.org/abs/1206.6272}{{\ttfamily arXiv:1206.6272 [hep-th]}}.

\bibitem{Eynard:2015aea}
B.~Eynard, T.~Kimura, and S.~Ribault, ``{Random matrices},''
  \href{http://arxiv.org/abs/1510.04430}{{\ttfamily arXiv:1510.04430
  [math-ph]}}.

\bibitem{MAKEENKO1991574}
Y.~Makeenko, A.~Marshakov, A.~Mironov, and A.~Morozov, ``Continuum versus
  discrete virasoro in one-matrix models,''
  \href{http://dx.doi.org/https://doi.org/10.1016/0550-3213(91)90379-C}{{\em
  Nuclear Physics B} {\bfseries 356} no.~3, (1991) 574 -- 628}.
  \url{http://www.sciencedirect.com/science/article/pii/055032139190379C}.

\bibitem{MORRIS1991703}
T.~Morris, ``Chequered surfaces and complex matrices,''
  \href{http://dx.doi.org/https://doi.org/10.1016/0550-3213(91)90383-9}{{\em
  Nuclear Physics B} {\bfseries 356} no.~3, (1991) 703 -- 728}.
  \url{http://www.sciencedirect.com/science/article/pii/0550321391903839}.

\bibitem{Livan_2018}
G.~Livan, M.~Novaes, and P.~Vivo, ``{Introduction to Random Matrices},''
  \href{http://dx.doi.org/10.1007/978-3-319-70885-0}{{\em SpringerBriefs in
  Mathematical Physics} (2018) }.
  \url{http://dx.doi.org/10.1007/978-3-319-70885-0}.

\bibitem{Cordova:2016jlu}
C.~Cordova, B.~Heidenreich, A.~Popolitov, and S.~Shakirov, ``{Orbifolds and
  Exact Solutions of Strongly-Coupled Matrix Models},''
  \href{http://dx.doi.org/10.1007/s00220-017-3072-x}{{\em Commun. Math. Phys.}
  {\bfseries 361} no.~3, (2018) 1235--1274},
\href{http://arxiv.org/abs/1611.03142}{{\ttfamily arXiv:1611.03142 [hep-th]}}.

\bibitem{Morozov_2019}
A.~Morozov, ``{On W-representations of $\beta$- and $q,t$-deformed matrix
  models},'' \href{http://dx.doi.org/10.1016/j.physletb.2019.03.047}{{\em
  Physics Letters B} {\bfseries 792} (May, 2019) 205–213}.
  \url{http://dx.doi.org/10.1016/j.physletb.2019.03.047}.

\bibitem{Alexandrov:2003pj}
A.~Alexandrov, A.~Mironov, and A.~Morozov, ``{Partition functions of matrix
  models as the first special functions of string theory. 1. Finite size
  Hermitean one matrix model},''
  \href{http://dx.doi.org/10.1142/S0217751X04018245}{{\em Int. J. Mod. Phys. A}
  {\bfseries 19} (2004) 4127--4165},
  \href{http://arxiv.org/abs/hep-th/0310113}{{\ttfamily arXiv:hep-th/0310113}}.

\bibitem{Alexandrov:2004ed}
A.~Alexandrov, A.~Mironov, and A.~Morozov, ``{Unified description of
  correlators in non-Gaussian phases of Hermitean matrix model},''
  \href{http://dx.doi.org/10.1142/S0217751X06029375}{{\em Int. J. Mod. Phys. A}
  {\bfseries 21} (2006) 2481--2518},
  \href{http://arxiv.org/abs/hep-th/0412099}{{\ttfamily arXiv:hep-th/0412099}}.

\bibitem{Alexandrov:2004ud}
A.~Alexandrov, A.~Mironov, and A.~Morozov, ``{Solving Virasoro constraints in
  matrix models},'' \href{http://dx.doi.org/10.1002/prop.200410212}{{\em
  Fortsch. Phys.} {\bfseries 53} (2005) 512--521},
  \href{http://arxiv.org/abs/hep-th/0412205}{{\ttfamily arXiv:hep-th/0412205}}.

\bibitem{Dimofte:2011py}
T.~Dimofte, D.~Gaiotto, and S.~Gukov, ``{3-Manifolds and 3d Indices},''
  \href{http://dx.doi.org/10.4310/ATMP.2013.v17.n5.a3}{{\em Adv. Theor. Math.
  Phys.} {\bfseries 17} no.~5, (2013) 975--1076},
  \href{http://arxiv.org/abs/1112.5179}{{\ttfamily arXiv:1112.5179 [hep-th]}}.

\bibitem{Kapustin:2009kz}
A.~Kapustin, B.~Willett, and I.~Yaakov, ``{Exact Results for Wilson Loops in
  Superconformal Chern-Simons Theories with Matter},''
  \href{http://dx.doi.org/10.1007/JHEP03(2010)089}{{\em JHEP} {\bfseries 03}
  (2010) 089}, \href{http://arxiv.org/abs/0909.4559}{{\ttfamily arXiv:0909.4559
  [hep-th]}}.

\bibitem{Hama:2010av}
N.~Hama, K.~Hosomichi, and S.~Lee, ``{Notes on SUSY Gauge Theories on
  Three-Sphere},'' \href{http://dx.doi.org/10.1007/JHEP03(2011)127}{{\em JHEP}
  {\bfseries 03} (2011) 127},
\href{http://arxiv.org/abs/1012.3512}{{\ttfamily arXiv:1012.3512 [hep-th]}}.

\bibitem{Alday:2013lba}
L.~F. Alday, D.~Martelli, P.~Richmond, and J.~Sparks, ``{Localization on
  Three-Manifolds},'' \href{http://dx.doi.org/10.1007/JHEP10(2013)095}{{\em
  JHEP} {\bfseries 10} (2013) 095},
  \href{http://arxiv.org/abs/1307.6848}{{\ttfamily arXiv:1307.6848 [hep-th]}}.

\bibitem{Pasquetti:2011fj}
S.~Pasquetti, ``{Factorisation of N = 2 Theories on the Squashed 3-Sphere},''
  \href{http://dx.doi.org/10.1007/JHEP04(2012)120}{{\em JHEP} {\bfseries 04}
  (2012) 120},
\href{http://arxiv.org/abs/1111.6905}{{\ttfamily arXiv:1111.6905 [hep-th]}}.

\bibitem{Amariti:2018wht}
A.~Amariti and L.~Cassia, ``{USp(2N$_{c}$) SQCD$_{3}$ with antisymmetric:
  dualities and symmetry enhancements},''
  \href{http://dx.doi.org/10.1007/JHEP02(2019)013}{{\em JHEP} {\bfseries 02}
  (2019) 013}, \href{http://arxiv.org/abs/1809.03796}{{\ttfamily
  arXiv:1809.03796 [hep-th]}}.

\bibitem{vdBult:2007}
F.~J. van-de Bult, ``Hyperbolic hypergeometric functions,''.
  \url{https://hdl.handle.net/11245/1.275741}.

\bibitem{Dijkgraaf:2002fc}
R.~Dijkgraaf and C.~Vafa, ``{Matrix models, topological strings, and
  supersymmetric gauge theories},''
  \href{http://dx.doi.org/10.1016/S0550-3213(02)00766-6}{{\em Nucl. Phys. B}
  {\bfseries 644} (2002) 3--20},
  \href{http://arxiv.org/abs/hep-th/0206255}{{\ttfamily arXiv:hep-th/0206255}}.

\bibitem{Dijkgraaf:2002vw}
R.~Dijkgraaf and C.~Vafa, ``{On geometry and matrix models},''
  \href{http://dx.doi.org/10.1016/S0550-3213(02)00764-2}{{\em Nucl. Phys. B}
  {\bfseries 644} (2002) 21--39},
  \href{http://arxiv.org/abs/hep-th/0207106}{{\ttfamily arXiv:hep-th/0207106}}.

\bibitem{Nazzal:2018brc}
B.~Nazzal and S.~S. Razamat, ``{Surface Defects in E-String Compactifications
  and the van Diejen Model},''
  \href{http://dx.doi.org/10.3842/SIGMA.2018.036}{{\em SIGMA} {\bfseries 14}
  (2018) 036}, \href{http://arxiv.org/abs/1801.00960}{{\ttfamily
  arXiv:1801.00960 [hep-th]}}.

\bibitem{Razamat:2018zel}
S.~S. Razamat, ``{Flavored surface defects in 4d $\mathcal{N}=1$ SCFTs},''
  \href{http://dx.doi.org/10.1007/s11005-018-01145-9}{{\em Lett. Math. Phys.}
  {\bfseries 109} no.~6, (2019) 1377--1395},
  \href{http://arxiv.org/abs/1808.09509}{{\ttfamily arXiv:1808.09509
  [hep-th]}}.

\bibitem{Faddeev:1993rs}
L.~Faddeev and R.~Kashaev, ``{Quantum Dilogarithm},''
  \href{http://dx.doi.org/10.1142/S0217732394000447}{{\em Mod. Phys. Lett. A}
  {\bfseries 9} (1994) 427--434},
  \href{http://arxiv.org/abs/hep-th/9310070}{{\ttfamily arXiv:hep-th/9310070}}.

\bibitem{Kurokawa:2003}
N.~{Kurokawa} and S.~{Koyama}, ``{Multiple sine functions.},'' {\em {Forum
  Math.}} {\bfseries 15} no.~6, (2003) 839--876.

\bibitem{Narukawa:2003}
A.~Narukawa, ``The modular properties and the integral representations of the
  multiple elliptic gamma functions,'' {\em ArXiv Mathematics e-prints} (Jun,
  2003) , \href{http://arxiv.org/abs/math/0306164}{{\ttfamily math/0306164}}.

\bibitem{Macdonald}
I.~G. Macdonald, {\em Symmetric Functions and Hall Polynomials}.
\newblock 01, 1995.
\newblock pp. 373-376.

\bibitem{Shiraishi:1995rp}
J.~Shiraishi, H.~Kubo, H.~Awata, and S.~Odake, ``{A Quantum deformation of the
  Virasoro algebra and the Macdonald symmetric functions},''
  \href{http://dx.doi.org/10.1007/BF00398297}{{\em Lett. Math. Phys.}
  {\bfseries 38} (1996) 33--51},
\href{http://arxiv.org/abs/q-alg/9507034}{{\ttfamily arXiv:q-alg/9507034
  [q-alg]}}.

\bibitem{Rains:2006dfy}
E.~M. Rains, ``{Limits of elliptic hypergeometric integrals},''
  \href{http://dx.doi.org/10.1007/s11139-007-9055-3}{{\em Ramanujan J.}
  {\bfseries 18} no.~3, (2007) 257--306},
  \href{http://arxiv.org/abs/math/0607093}{{\ttfamily arXiv:math/0607093}}.

\bibitem{Willett:2011gp}
B.~Willett and I.~Yaakov, ``{N=2 Dualities and Z Extremization in Three
  Dimensions},'' \href{http://arxiv.org/abs/1104.0487}{{\ttfamily
  arXiv:1104.0487 [hep-th]}}.

\end{thebibliography}
\end{document}